\documentclass[useAMS,usenatbib]{mnras}

\usepackage{graphicx,amsmath,amssymb,hyperref}
\usepackage[T1]{fontenc}
\usepackage{aecompl}
\usepackage{subcaption}
\usepackage{amsmath}
\usepackage{picture}
\title[Introducing 21SSD]{21SSD: a public database of simulated 21-cm signals from the epoch of reionization }
\author[B. Semelin et al.]{B. Semelin$^{1}$\thanks{e-mail:
benoit.semelin@obspm.fr}, E. Eames$^{1}$\thanks{e-mail:evan.eames@obspm.fr}, F. Bolgar$^{1}$, M. Caillat$^{1}$.\\
$^{1}$ Sorbonne Universit\'es, UPMC, LERMA, Observatoire de Paris, PSL research university, CNRS, F-75014, Paris, France}
\begin{document}

\date{Accepted XXX. Received XXX; in original form XXX}

\pagerange{\pageref{firstpage}--\pageref{lastpage}} 

\maketitle

\label{firstpage}

\begin{abstract}
The 21-cm signal from the Epoch of Reionization (EoR) is expected to be detected in the next few years, either with existing instruments or by the upcoming SKA and HERA projects. In this context there is a pressing need for publicly available high-quality templates covering a wide range of possible signals. These are needed both for end-to-end simulations of the up-coming instruments, as well as to develop signal analysis methods. In this work we present such a set of templates, publicly available, for download at \href{https://21ssd.obspm.fr/}{21ssd.obspm.fr}. The database contains 21-cm brightness temperature lightcones at high and low resolution, and several derived statistical quantities for 45 models spanning our choice of 3D parameter space. These data are the result of fully coupled radiative hydrodynamic high resolution ($1024^3$) simulations performed with the LICORICE code. Both X-ray and Lyman line transfer is performed to account for heating and Wouthuysen-Field coupling fluctuations. We also present a first  exploitation of the data using the power spectrum and the Pixel Distribution Function (PDF) as  functions of redshifts, computed from lightcone data. We analyse how these two quantities
behave when varying the model parameters while taking into account the thermal noise expected of a typical SKA survey. Finally, we show that the power spectrum and the PDF have different -- and to some extent complementary -- abilities to distinguish between different models. This opens the door to formulating an optimal sampling of the parameter space, dependant on the chosen diagnostics.   

\end{abstract}


\section{Introduction}

In the coming decade our knowledge of the history of the universe during the period of primordial galaxy formation is expected to make a great leap forward. One of the most promising observational probes is the $21$-cm signal emitted in neutral regions of the intergalactic medium (IGM) during the Epoch of Reionization (EoR), between redshifts $\sim 27$ and $\sim 6$. The angular fluctuations and redshift evolution of this signal encode a wealth of information on the nature and formation history of primordial sources because of sensitivity to Lyman-$\alpha$, ionizing UV, and X-ray photon production through various processes \citep[for a review, see][]{Furlanetto06}. As the signal is redshifted to different wavelengths depending on the redshift of the emitting region, we can, in principle, build a full tomography of the IGM between $z \sim 27$ and $z \sim 6$ by observing in the $50-200$ MHz band. However, the thermal noise due primarily to the sky brightness is daunting, especially at low frequencies, and an SKA-like
effective collecting area will be realistically required to build a tomography with a good signal-to-noise ratio,  even at moderate redshift ($< 12$) and resolution ($\sim 5^{\prime}$) \citep{Mellema13}. This is why the instruments currently in operation focus on obtaining a statistical detection of the signal through its power spectrum, which benefits from a much better signal-to-noise ratio. 

To date, only upper limits on the power spectrum of the signal have been published. At redshift $z \sim 8$, projects using the GMRT and MWA have both published upper limits in the $(200\, \mathrm{mK})^2$ range \citep{Paciga13, Dillon15}. LOFAR has a $(80\, \mathrm{mK})^2$ upper limit at $z \sim10$ and $k=0.05 $ h.Mpc$^{-1}$ limited by systematics \citep{Patil17},   while observations with
 PAPER yielded a $(22\, \mathrm{mK})^2$ upper limit at $z=8.4$ and $k=0.3 $ h.Mpc$^{-1}$ \citep{Ali15}. From this last result it is possible to exclude models of the EoR with very low X-ray production \citep{Pober15}. The project at MWA also carried out pioneering observations at $z \sim 16$ \citep{Ewall-Wice16}. These upper limits will hopefully soon be followed by actual detections, either with current instruments or upcoming ones such as the SKA or HERA. With this in mind, we can appreciate the urgency of developing methods allowing us to extract  astrophysical information from observations in a systematic way.
 
 The first step is to define a set of astrophysical parameters that determine the signal. The definition of this set is neither straightforward, nor unique. Some parameters are obvious choices, for example the star formation efficiency. But even in this case different parameterization can be defined (e.g. a gas conversion time scale or a fraction of the host halo mass) and different levels of modelling can be adopted (redshift evolution of the parameters, influence of metallicity, etc.). Other parameters are closely
 connected to the specific modelling method being used. For example, the feedback of the ionization field on the star formation rate in the smaller halos is automatically included in radiative hydrodynamic simulations (with a level of realism limited by the finite resolution) but will appear as a parameter in simulations where radiative transfer is performed in post-processing, as well as in semi-numerical simulations. Various parameterizations have already been suggested \citep{Greig15,Cohen16,Greig17}, and in this work we will re-use some existing parameters while defining some new ones. Since exploring the parameter space is usually a computationally intensive task, it will be important that some level of convergence is reached in the coming years concerning the definitions, so that a synergy between the efforts of different teams can emerge. The SKA EoR science working group is currently aiming for this convergence.
 
 Once a parameter space is defined, the goal is to use observations to put constraints on acceptable values of said parameters. Methods used to constrain cosmological parameters from CMB data can, to some extent, be applied to the $21$-cm signal. The requirement is that the forward modelling (i.e. deriving the signal for a given set of parameter values) can be performed at limited computational cost, as the methods used in CMB observations usually require many instances of forward modelling. An  example of such a method is that of the Markov Chain Monte Carlo, which has been applied in combination with the 21cmFast semi-numerical code \citep{Greig15}, or that of a neural network based emulator \citep{Kern17}. Or one can use the Fisher matrix formalism, combined with  the above mentioned semi-numerical code \citep{Pober14,Liu16,Ewall-Wice16b}. On the other hand, one may want to rely on more accurate  \citep{Zahn11}, but  computationally more expensive, full numerical simulations. In this case, one must use methods that can handle a sparse parameter space sampling. Such an approach is used in \citet{Shimabukuro17} in which a neural network is trained on a sparse sampling of a choice of 3-dimensional parameter space to perform backward modelling: that is, derive the parameter values for a given signal (\citet{Shimabukuro17} actually also use 21cmFast, but aim to estimate the validity of the approach, which can then be used on full numerical simulations). 
 
 Even with a sparse sampling of the parameter space, full simulations rapidly become computationally expensive as the dimension of the parameter space grows. Indeed, simulations which cover a sufficient cosmological volume to mitigate cosmic variance \citep[see][]{Iliev14} and resolve halos with masses  not much larger than the atomic cooling threshold (i.e. $10^8$ M$_\odot$) require billions of resolution elements. A single such simulation can easily reach $10^6$ computing hours if gas dynamics and accurate radiative transfer are included. This is the main motivation behind this work: to provide the community with high quality simulated signals resulting from the sparse sampling (45 points) of a 3D parameter space. Although, in  future higher resolution simulations as well as other parameters will have to be explored, this is a solid first step. The data presented in this work were produced in about $3. 10^6$ CPU hours  and can now be used by other teams to test existing or new methods of deriving parameter constraints.
 
 The database presented in this work can also be used as a collection of templates spanning a wide range of (but not all) possible signals. This could be useful for end-to-end simulation of upcoming instruments such as the SKA. A third possible
 use is to cross-check with fast semi-numerical codes, and possibly tune them to improve prediction
 agreement. For example, another publicly available database has been produced with 21cmFast \citep{Mesinger16} which contains simulations quite complementary to 21SSD: they cover a much larger cosmological volume (a cube with side roughly 5 times larger, close to the size of a SKA survey) with the same number of resolution elements, and thus lower spatial resolution. While both types of simulations are useful, they would be even more so if one could ensure that they result from convergent numerical methods. In the same spirit, fast simulation codes tuned to replicate the results of full simulations on a sizeable region of the parameter space could be used to obtain a much finer sampling.
 
 Let us finally briefly emphasize the salient features of the simulations presented in this work (see section 2 for more details). These are fully coupled radiative hydrodynamic simulations performed in a $200$ Mpc.h$^{-1}$ box with $1024^3$ particles, allowing us to resolve halos of $\sim 10^{10}$ M$_\odot$. Higher resolution simulations in smaller volumes have shown that halos with mas $< 10^9$ M$_\odot$ have their star formation efficiency decreased by radiative feedback \citep[e.g.][]{Ocvirk16}. Thus we model a fair faction of the radiation emitted in atomic cooling halos. It is important to mention that the ionizing UV and X-ray radiative transfer is performed on an adaptive mesh with a spatial resolution similar to that used for dynamics ($\sim 5$ cKpc). We also perform Lyman line radiative transfer to account for the Wouthuysen-Field effect so that we are able to compute the 21-cm signal at all stages of reionization, but it is performed as a post-processing step on a fixed grid. The combination of rich physical modelling, reasonably good resolution, and large simulation volumes place our simulations among the forerunners of the field of 21-cm signal predictions. As well, the exploration of chosen the parameter space should make the data interesting to the community.  
 
 In section 2, we present the numerical methods and describe the simulation setup. Section 3 details the data publicly available in the database and draws some first conclusions. In section 4, we present a first exploitation of the database, and finally present our conclusions in section 5.

\section{Numerical model}
\subsection{The LICORICE code}
The data in 21SSD were created using the LICORICE code. The incremental development of this code is described in a number of papers \citep{Semelin02, Semelin07, Baek09, Iliev09, Baek10,  Vonlanthen11, Semelin16}. Here we will summarize its main features and describe minor new ones. LICORICE uses a Tree+SPH method for implementing gravity and hydrodynamics \citep{Semelin02}. Radiative
transfer in the ionizing UV and X-ray frequencies is coupled to the dynamics and implemented using Monte Carlo ray-tracing on an adaptive
grid, in turn derived from the tree used for calculating gravity \citep{Baek09, Iliev09, Baek10}. The ionization of both H and He is implemented, although only H is used in 21SSD. The energy and ionization equations are integrated with individual adaptive time-steps for each particle, sub-cycling within dynamical time steps. The correct value for the speed of light is used when propagating the Monte Carlo photon packets. To compute the spin temperature of hydrogen, the local Lyman alpha flux is required. LICORICE computes the transfer in the resonant Lyman lines on a fixed grid, post-processing the radiative hydrodynamic simulations \citep{Semelin07,Vonlanthen11}. The radiative hydrodynamic part of LICORICE benefits from double MPI+OpenMP parallelization \citep[briefly described in][]{Semelin16}, while the Lyman line transfer is OpenMP parallelized with a simple MPI overlay (no domain decomposition).

The main new feature of LICORICE, introduced for 21SSD, is the implementation of a hard X-ray contribution from sources such as X-ray binaries. The spectrum is taken from \citet{Fragos13}, with a varying luminosity (see section 3.2). The difficulty is that during the EoR the mean free path at energies above $1$ keV is large: a considerable fraction of the energy is not absorbed before the end of the EoR. With a Monte Carlo algorithm, this would require  tracking a tremendous number of photons by the end of the simulation. Thus, we tag photons that have travelled a distance larger than the box size as \textit{background }photons. Whenever the number of background photons reach a fixed threshold, we kill a fraction of them, redistributing their energy content amongst the others. Using a large number of photons (in the $10^9$ range), we trust that the global spectral properties of the background is unchanged during this operation (local properties are not relevant since these are background photons).

\subsection{Simulation setup}
We now describe the features common to all simulations in 21SSD. The simulations cover a $200$ h$^{-1}$.Mpc cube and include $1024^3$ particles, half for the gas and half for dark matter. In the adopted cosmology 
\citep[$H_0=67.8$km.s$^{-1}$, $\Omega_m=0.308$, $\Omega_\Lambda=0.692$, $\Omega_b=0.0484$, $\sigma_8=0.8149$ and $n_s=0.968$ ][]{Planck15} this corresponds, respectively, to masses of $2.9\,\, 10^8$ M$_{\odot}$ and $1.6\,\, 10^9$ M$_\odot$. The initial conditions were generated at $z=100$ using second order Lagrangian Perturbation Theory via the Music package \citep{Hahn11}. The dynamics are computed using a fixed $1$ Myr time step ($0.33$ Myr at expansion factor $a < 0.03$) and the
gravitational  softening is $5$ ckpc.
The implementation of star formation is the same as in \citet{Semelin16}. The specific values of the parameters (overdensity threshold, gas conversion time scale and escape fraction) are discussed in section \ref{sec:calibration}. The only feedback from star formation comes from photo-heating. No sub-grid kinetic nor thermal feedback is active. 

For every dynamical time step, each particle containing a stellar fraction emits $2 \times \mathrm{min}(10^5,5 \,{10^7 \over \mathrm{nb\, of\, sources}})$ photon packets, half as ionizing UV, half as X-ray. By the end of each simulation, we are propagating $\sim 15\,\, 10^9$ photon packets. This ensures that, on average, each cell is crossed by $\sim 100$ photons for each time step. The UV photon frequency is chosen by a Monte Carlo sampling of the spectrum resulting from a Salpeter initial mass function truncated at $1.6\, \mathrm{M}_\odot$ and $120\,\, \mathrm{M}_\odot$ \citep{Baek10}. The X-ray spectrum is described in section \ref{sec:parameters}. Each radiative hydrodynamic simulation was typically run on  $4000$ cores, requiring $150\,000$ computing hours.

Computing the $21$-cm brightness temperature during the Cosmic Dawn, when the Wouthuysen-Field effect does not saturate, requires us to evaluate the local Lyman-$\alpha$ flux. This is performed while post-precessing  the dynamical simulations using the same method as in previous works \citep[e.g.][]{Semelin07,Vonlanthen11, Semelin16}. We used a $512^3$ fixed grid, emitting $4\,\, 10^8$ photons every $10^7$ years (the time interval between two snapshots of the dynamical simulation). As, for now, all simulations share a very similar star formation and ionization history (indeed the impact of varying the model parameters  on these two quantities is very small), we ran the Lyman-$\alpha$ simulation only once (see also section \ref{sec:parameters}). If, in future, we extend the database in such a way that the local star formation history and/or ionization state of the IGM are significantly changed, then we will have to re-run the Lyman-$\alpha$ simulation.

\section{Description of the database}
\subsection{Calibration choices}
\label{sec:calibration}
We chose, for this first release of the database, to calibrate the simulations using several observational constraints. Since these constraints are all connected to the production of ionizing photons to some degree, and since the star formation history and escape fraction vary little to none across the parameter space region we explore, \textit{all} the simulations in the database match the constraints to the same degree of accuracy. In future extensions of the database, some of these constraints will be relaxed within their error bars, and parameters such as the escape fraction will be varied.

\begin{figure}
\includegraphics[width=0.49\textwidth]{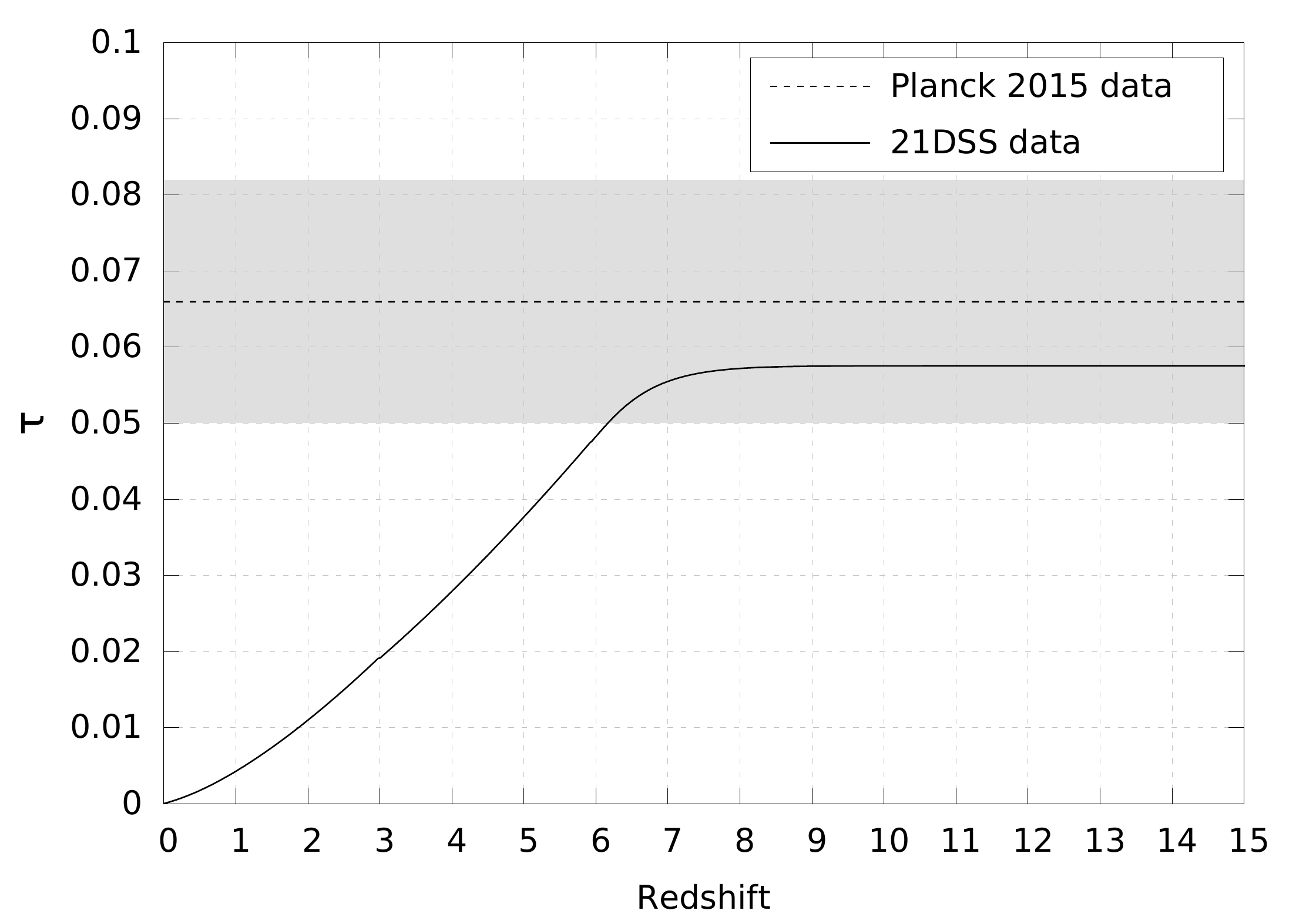}
\caption{Average Thomson scattering optical depth computed out to redshift $z$. The simulations are used at $z >6$ and we assume identical first Hydrogen and Helium ionization, as well as instantaneous second Helium ionization at $z=3$. The history does not depend on the model since they have almost identical ionization histories.}
\label{thomson}
\end{figure}

\subsubsection{Thomson scattering optical depth and reionization history}
\begin{figure}
\includegraphics[width=0.49\textwidth]{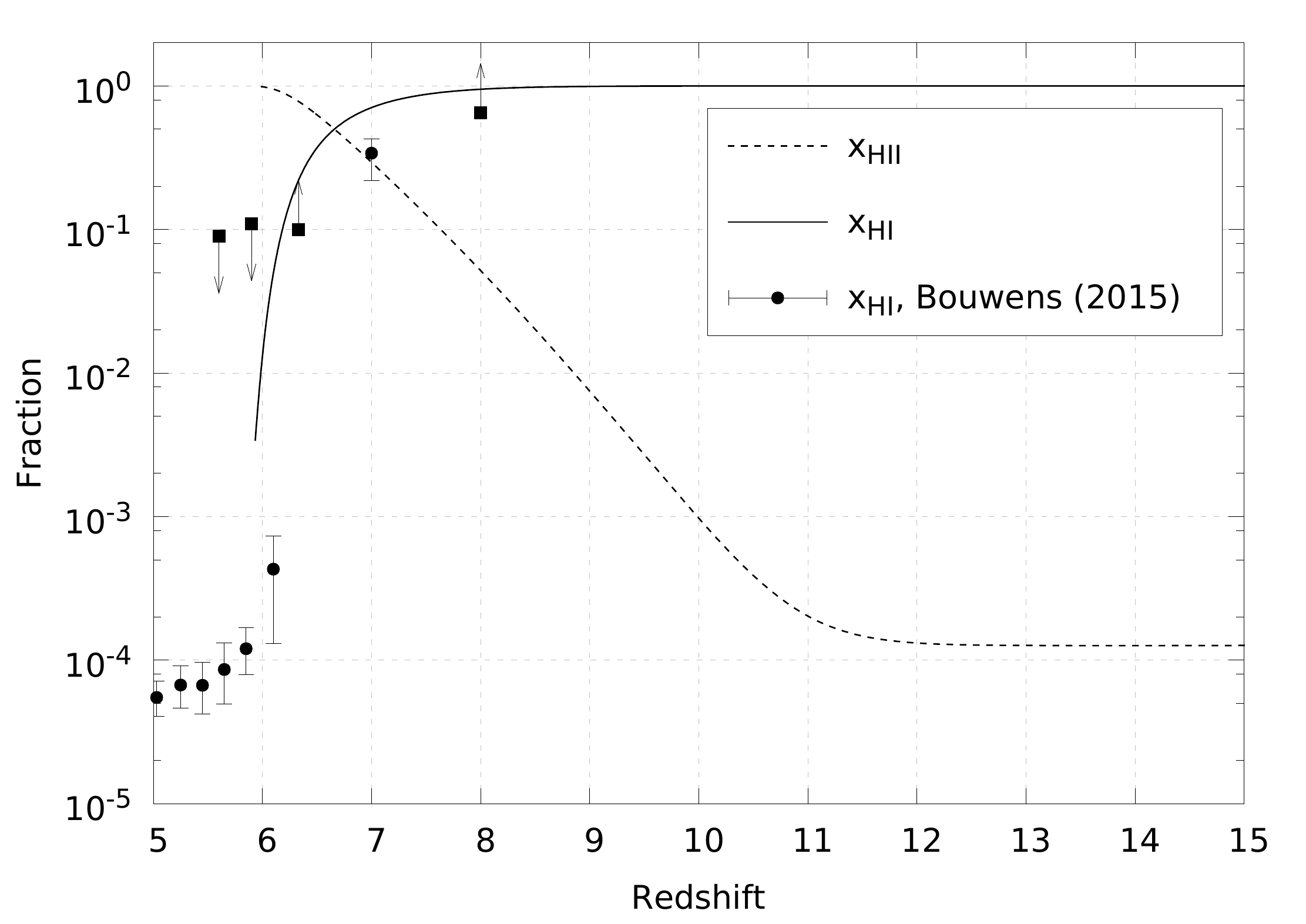}
\caption{Evolution over redshift of the volume averaged ionized and neutral fraction, compared with observational data. All models share a near identical history (differing only in X-ray production).}
\label{ion_hist}
\end{figure}

There are now a number of observational constraints on the history of reionization. First, interpretations of CMB observations yield a constraint on the Thomson scattering optical depth $\tau$, which translates into an integral constraint on the reionization history. Fig. \ref{thomson} shows the increase of $\tau$ as a function of $z$ in the simulations (assuming an instantaneous second He ionization at $z=3$) and the corresponding value measured by \citet{Planck15}. Our simulations fall within $0.5 \,\, \sigma$ of the observations. There are now a number of constraints on the average ionized fraction at specific redshifts derived from observations. Some of these constraints are reviewed in \citet{Bouwens15a}. In fig. \ref{ion_hist} we plot the ionization history in 21SSD against observational data of the Gunn-Peterson effect in QSO spectra \citep{Fan06}, the statistics of dark gaps in QSO spectra \citep{McGreer15}, Lyman-$\alpha$ damping wings in QSO spectra \citep{Schroeder13} and Lyman-$\alpha$ emission in galaxies \citep{Schenker14}, as summarized in \citet{Bouwens15a}. Our reionization history fits the observational constraints reasonably well.

\begin{figure}
\includegraphics[width=0.49\textwidth]{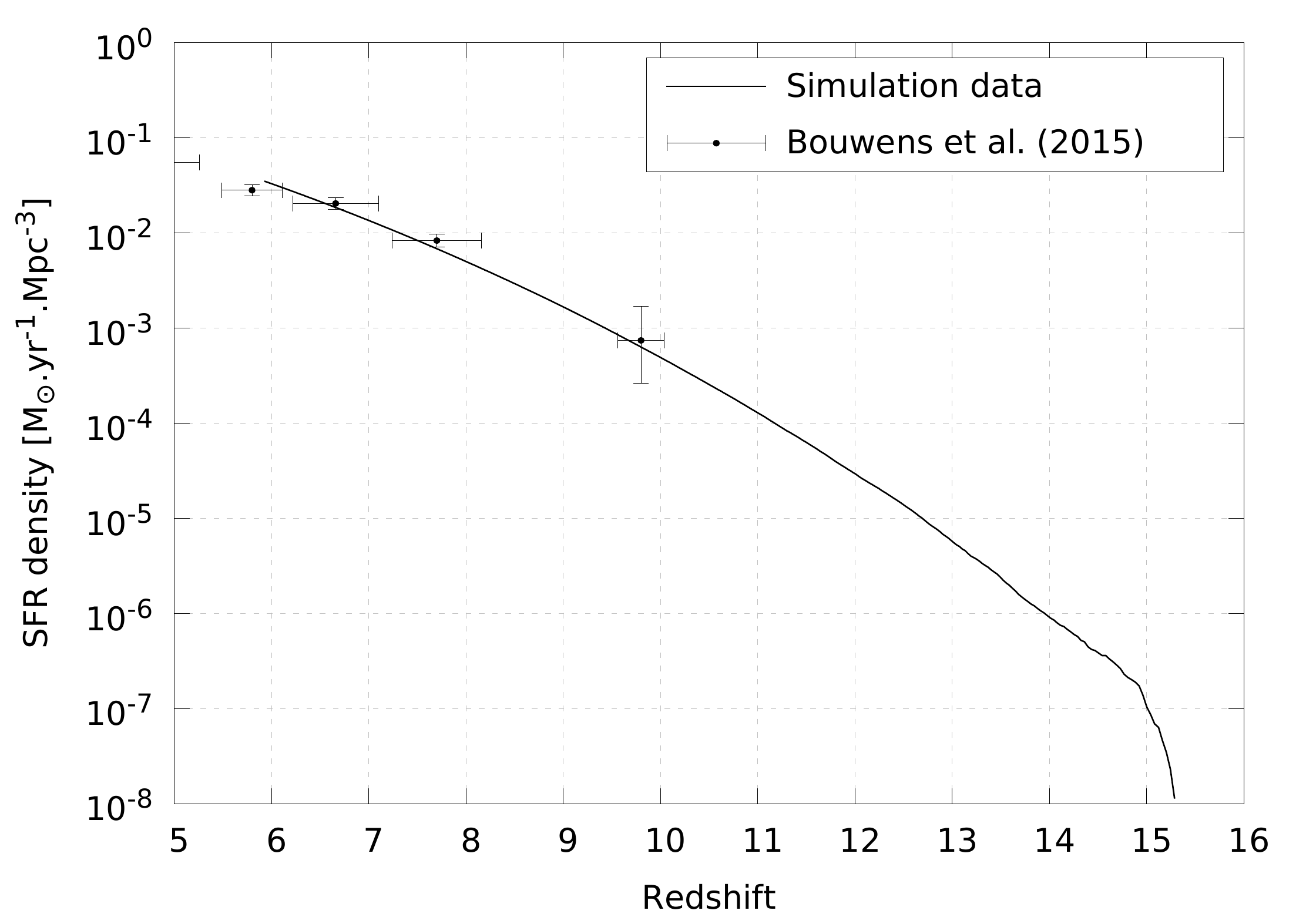}
\caption{History of the average SFR density in the simulations, compared to observational data. The match was achieved by tuning the simulation parameters. See main text for the computation of the redshift value and error bars for the observational data points. }
\label{star_hist}
\end{figure}

\subsubsection{Cosmic star formation history}

Taking into account the available observational constraints on the cosmic star formation history allows us to break the degeneracy between the escape fraction of ionizing photons and the star formation efficiency (which exists if only the global ionization history is considered). In  fig. \ref{star_hist} we plot the evolution of the SFR density as a function of redshift in the simulations, together with observational constraints derived by \citet{Bouwens15b}. The mean-redshift values and horizontal error bars of the observational points were not computed from the definition of the redshift bins, but instead from the actual mean and scatter of the populations in each bin. To achieved this match we tuned only the star formation efficiency in the simulations.
\subsection{Astrophysical parameters}
\label{sec:parameters}
Modelling the $21$-cm signal from the EoR, even with detailed numerical simulations, involves a number of parameters that encapsulate the action of physical processes beyond the scope of the simulation (usually because they operate on scales outside of the range covered by the simulations). Some of these parameters can be constrained by observations as in the previous section while others are relatively free, such as the efficiency of X-ray production at high redshift. Let us review the parameters kept constant in this first version of the database (possibly to be varied later on) and those that were varied.
\subsubsection{Unvaried parameters}

As mentioned in section \ref{sec:calibration}, the parameters describing star formation were calibrated to match observational data, and therefore remain unchanged throughout the simulations. Star formation is triggered in gas particles with a comoving overdensity above $100$. We use a Schmidt law with exponent $1$: ${d\rho_s \over dt} = c_{\mathrm{eff}} \rho_g$, where $\rho_g$ is the gas density, $\rho_s$ is the star density and $ c_{\mathrm{eff}}$ is an efficiency parameter. In the case of an exponent equal to $1$, $c_{\mathrm{eff}}$ is the inverse of the gas conversion time scale. This time scale is set to $2$ Gyr in the simulations. Note that the physical meaning of this value should not be overemphasized: it was set such that it satisfied the observational SFR density constraints and, in fact, depends strongly on the chosen density threshold that, in turn, is related to the mass resolution of the simulation. Only the photo-heating feedback, intrinsic to radiative hydrodynamics, is effective: no kinetic nor thermal feedback from SN or AGN is included. The Initial Mass Function (IMF) of the star population is chosen to be a Salpeter IMF with lower and upper mass cutoffs at $1.6$ and $120$ M$_\odot$ respectively. The resulting emissivity and spectral shape in the Lyman lines and ionizing UV ranges are then computed using the procedure described in \citet{Baek10}. The emissivity in the range of the Lyman lines is, in fact, varied as explained in the next section.

To satisfy the observational constraints on the ionization history, the escape fraction of ionizing UV photons was fixed at $f_{\mathrm{esc}}=0.2$. Note that this quantity refers to the photons escaping from dense structures around the source and below the resolution of the simulation, not from halos: photons are explicitly propagated and absorbed in resolved structures within the halos. Thus, while $f_{\mathrm{esc}}=0.2$ does not depend on halo mass, different behaviour can occur in halos with different masses, especially towards the well-resolved high-mass end. 

Although the emissivity of hard and soft X-rays varies between simulations (see next section), the spectral shapes do not. Soft X-rays, for example those produced by AGNs, are modelled with a spectral index of $1.6$, a lower cutoff at $100$ eV, and an upper cutoff at $2$ KeV. Hard X-rays, due mainly to the contribution of X-ray binaries, are modelled using the spectral properties tabulated by \citet{Fragos13}.

\begin{figure}
\includegraphics[width=0.49\textwidth]{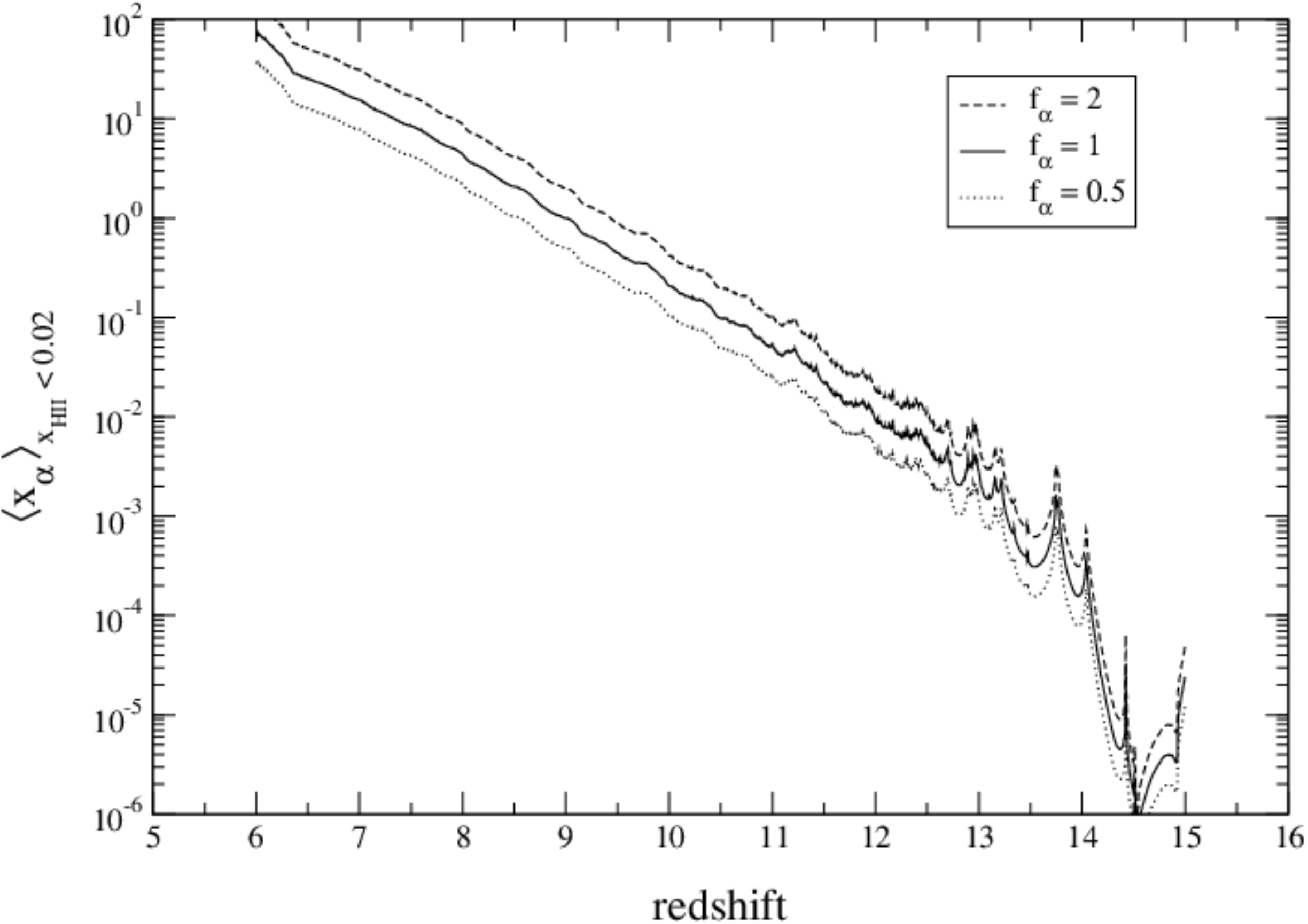}
\caption{History of the spatially averaged $x_\alpha$ coefficient in the neutral IGM  (defined as $x_{\mathrm{HII}}<0.02$), after multiplying by $f_\alpha$ but before taking the backreaction into account.  }
\label{mean_xa}
\end{figure}

\begin{figure}
\includegraphics[width=0.49\textwidth]{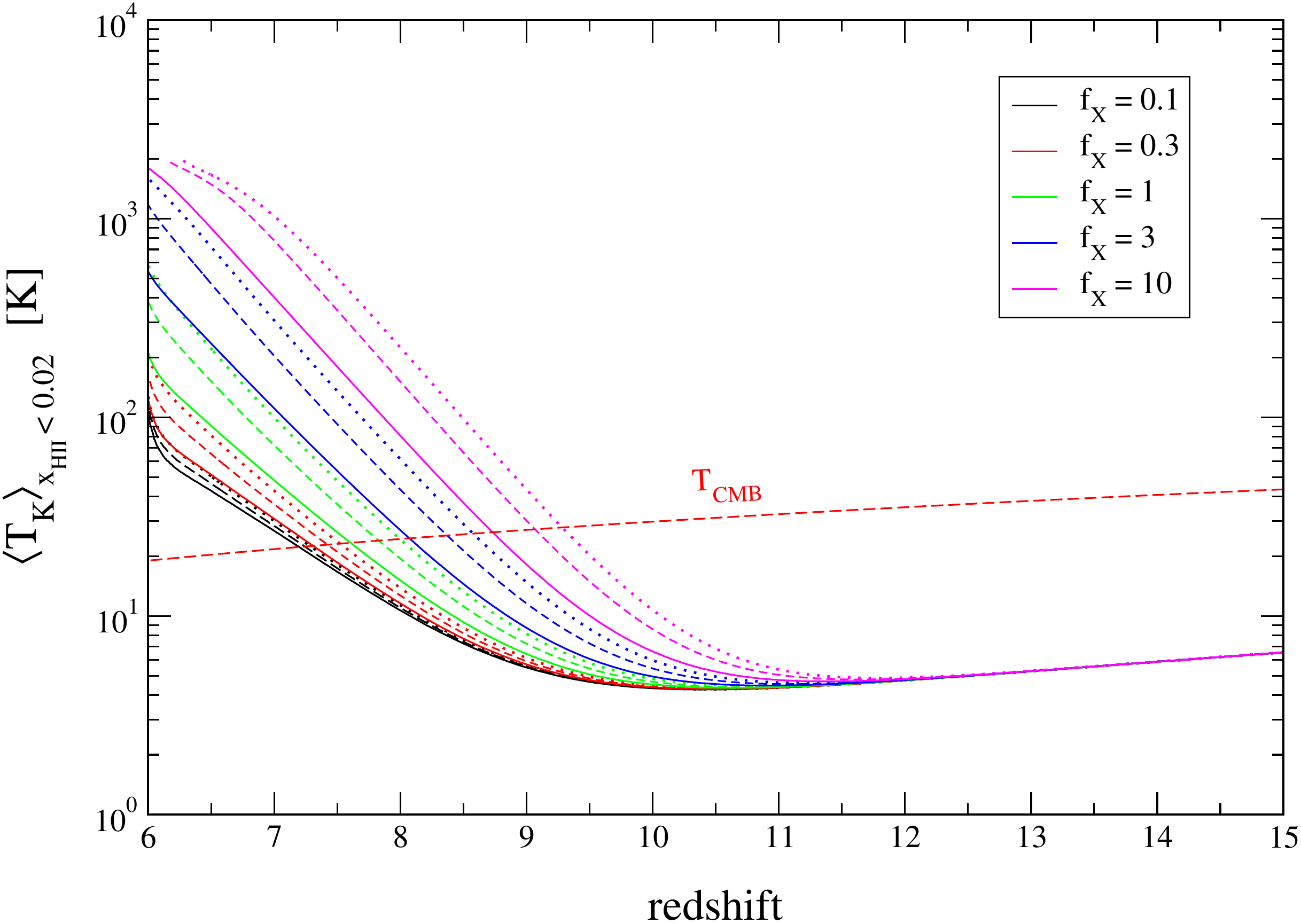}
\caption{Histories of the average kinetic temperature in the neutral gas (defined as $x_{\mathrm{HII}}<0.02$) for all the models. The solid lines correspond to models with $r_{H/S}=1$, dashed lines to $r_{H/S}=0.5$, and dotted lines to $r_{H/S}=0.0$. Colour encodes the value of $f_X$, as specified in the label. }
\label{mean_Tk}
\end{figure}

\subsubsection{A choice of 3-parameter space}
\label{sec:varied_param}

In this study, we chose to vary parameters related to the Wouthuysen-Field coupling and the X-ray heating of the neutral IGM. Such parameters crucially determine the duration and intensity of the era in which the $21$ cm signal can be seen in absorption against the CMB.

{\sl Lyman band emissivity: $f_\alpha$}

\noindent
The $21$ cm brightness temperature depends on the spin temperature (see eq. \ref{dtb_eq}), which, in turn, depends on the strength of the Wouthuysen-Field coupling \citep{Wouthuysen52,Field58} quantified by the the $x_\alpha$ coefficient. $x_\alpha$ can be computed from the local Lyman-$\alpha$ flux
$J_\alpha$ using the procedure described in \citet{Hirata06} that includes the backreaction of the gas on the shape of the spectrum near the Lyman-$\alpha$ line. To obtain the local Lyman-$\alpha$ flux, we perform the full radiative transfer computation in the Lyman lines \citep[see][]{Vonlanthen11}. Then,
the driving factor is the emissivity of the sources in the Lyman band. Since, in our modelling, we do not consider the impact of metal enrichment (e.g. changing IMF, absorption by dust), we use a constant luminosity emitted in the Lyman band. For more advanced modelling, it could vary in space and time self-consistently with metal enrichment. We use a more basic approach: we introduce an $f_\alpha$ coefficient to describe the Lyman band emissivity efficiency. Let us define the energy emitted between two frequencies by a stellar population representative of the IMF $\xi(M)$  during its lifetime as:

\begin{equation}
E(\nu_1,\nu_2)= \int_{\nu_1}^{\nu_2} \int_M \xi(M) L(M,\nu) T_{\mathrm{life}}(M) dM d\nu
\end{equation}

where $M$ is the stellar mass, $\nu$ the frequency, $T_{\mathrm{life}}(M)$ the lifetime of a star with mass $M$, and $L(M,\nu)$ the energy emitted per second per Hertz by a star of mass $M$ around frequency $\nu$. Then $f_\alpha$
is defined as:

\begin{equation}
f_\alpha={ E^{\mathrm{eff}}(\nu_\alpha,\nu_{\mathrm{limit}}) \over E(\nu_\alpha,\nu_{\mathrm{limit}}) }
\end{equation} 

where $\nu_\alpha$ is the Lyman-$\alpha$ frequency, $\nu_{\mathrm{limit}}$ is the Lyman limit frequency and $E^{\mathrm{eff}}$ is the energy effectively emitted in the simulation (rather than the theoretical one). We assumed the same spectral shape for the effective emission as for the theoretical one. For our purpose, we considered $f_\alpha=0.5, 1$, and $2$. This limited range is dictated by the fact that, at fixed ionizing emissivity, a substantial change in the IMF is required to significantly alter $f_\alpha$. Let us finally mention that varying this parameter in such a way effectively costs nothing in terms of computing time: since the Lyman line transfer has negligible feedback on the dynamics it is performed in post-processing, and the resulting radiation field is linear with the source emissivity. Thus, the Lyman line transfer simulation needs only to be performed once. If, in future, a new parameter is varied that affects the star formation history, the Lyman line transfer will have to be recomputed. Fig. \ref{mean_xa} shows the evolution of the mean $\langle x_\alpha \rangle$ value for all three choices of $f_\alpha$ (the average being restricted to neutral regions: $x_{\mathrm{HII}} < 0.02$). Note that the  $x_\alpha$ values are considered before the backreaction is applied \citep[see][]{Hirata06}. 

\begin{table}
\centering
\begin{tabular}{ll}
\hline
Parameter & Explored values \\
\hline
$f_\alpha$ & 0.5, 1., 2. \\
$f_X$ & 0.1, 0.3, 1., 3., 10. \\
$r_{H/S}$ & 0., 0.5, 1. \\
\hline
\label{tab:param}
\end{tabular}
 \caption{ The explored values for each parameters. Every combination is considered, hence 45 points in parameter space. The definition of the parameters is given in the main text (section \ref{sec:varied_param}).}
 \end{table}

{\sl X-ray emissivity: $f_X$}

\noindent
The X-ray heating of the neutral IGM is what drives the transition between the absorption and emission regimes for the $21$ cm signal.
The efficiency of X-ray production during the EoR is usually parametrized as:
\begin{equation}
L_X=3.4\,10^{40} f_X \left({\mathrm{SFR} \over 1\, {\mathrm{M_\odot.yr}^{-1}}}\right) \mathrm{erg.s^{-1}}
\end{equation}
where $L_X$ is the luminosity of a source, SFR the star formation rate, and $f_X$ an unknown correction factor between low and high redshift \citep{Furlanetto06d}. Every time we create a new source in the simulation we compute an equivalent steady state SFR using said source's mass and lifetime. Then, applying the above formula, we obtain its X-ray luminosity, using the same $f_X$ value throughout a single simulation. Since $f_X$ is highly uncertain we varied it across two orders of magnitude $f_X=0.1, 0.3, 1, 3$ and $10$.

{\sl X-ray hard-to-soft ratio: $r_{H/S}$}

It is known that different sources of X-rays (typically AGN vs X-ray binaries) have different spectra and heat the IGM with varying efficiencies \citep[e.g.][]{Fialkov14}. Indeed, sources with a hard spectrum will be less efficient since their X-ray photons will have mean free paths that increase as the cube of their energy. Thus, in addition to varying the X-ray  production efficiency between simulations, we also vary the ratio of energy emitted by AGN and X-ray binaries. 
If we define partial $f_X$ for AGN and X-ray binaries such that $f_X=f_X^{\mathrm{\scriptscriptstyle AGN}}+f_X^{\mathrm{\scriptscriptstyle XRB}}$, then:
\begin{equation}
r_{H/S}={ f_X^{\mathrm{\scriptscriptstyle XRB}} \over f_X}
\end{equation}

  This ratio takes the values $r_{H/S}=0$, $0.5$ and $1$. Fig. \ref{mean_Tk} shows the combined effect of varying $f_X$ and $r_{H/S}$  on the average kinetic temperature history of the gas regions with ionized fraction smaller than $0.02$. It shows that changing $r_{H/S}$ from $0$ to $1$ is more or less equivalent, in term of the average temperature, to decreasing $f_X$ by a factor of $3$. But of course, it is not that simple, and changing the $r_{H/S}$ value strongly alters the temperature fluctuations. This effect is not negated by adjusting $f_X$ in order to keep the average constant. 

\subsection{The 21-cm signal}
\subsubsection{Methodology}
The general formula for the differential brightness temperature against the CMB is:
\begin{multline}
\delta T_b \,=\, 27. \,\,(1-x_{\mathrm{H_{II}}}) \,(1+\delta)  \left( T_s - T_{\mathrm{cmb}} \over T_s \right) \left( 1 + {1 \over H(z)} {d v_{||} \over dr_{||}} \right)^{-1} \\ \times  \,\left( {1+z \over 10} \right)^{1 \over 2}   
\left({\Omega_b \over 0.044} {h \over 0.7} \right) \left({\Omega_m \over 0.27 }\right)^{1 \over 2} \,\,\,\mathrm{mK}
\label{dtb_eq}
\end{multline}
where $x_{\mathrm{H_{II}}}$ is the local ionized fraction of hydrogen, $\delta$ is the overdensity of the gas, $T_s$ is the local spin temperature of hydrogen, $T_{\mathrm{cmb}}$ the CMB temperature at that redshift, $H(z)$ the Hubble parameter, $d v_{||} \over dr_{||}$ the velocity gradient along the line of sight, $z$ the redshift, and the usual notation for cosmological parameters is used. Most contributions, including $T_s$, can be readily computed from the simulation data. $T_s$ is computed from the density, kinetic temperature, and local Lyman-$\alpha$ flux using the procedure described in \citet{Hirata06} that includes the backreaction of the gas on the spectrum. In the above formula, the velocity gradient along the line of sight can produce a spurious divergence. Instead of computing the velocity gradient, we follow the more robust method of moving the particles along the line of sight depending on their velocity, and then recomputing the density in the resulting redshift-space. This procedure was described in \citet{Semelin16}.

\begin{figure}
\includegraphics[width=8cm]{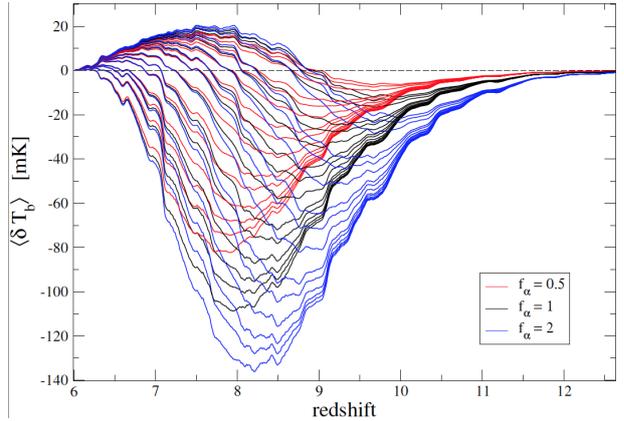}
\caption{Global $21$-cm signal computed from lightcone data for all $45$ models. Models are not identified individually, but each group of fixed $f_\alpha$ are ordered the same as in fig. \ref{mean_Tk} (strong to weak absorption, and weak to strong heating). This figure is intended to give an impression of the range of possible signals. Note that the small scale fluctuations consistent across models are caused by sample variance in the lightcone.}
\label{mean_dtb}
\end{figure}

\subsubsection{The range of predicted signals}

The explored parameter values are summarized in table 1. Since we consider every possible combination, we ultimately explore
45 points (i.e 45 models) in this 3-dimensional parameter space. The resulting global signal histories are presented in fig \ref{mean_dtb}. These were computed from the corresponding high resolution lightcones (see section \ref{sec:lightcones}) by averaging $\delta T_b$ in each of the 8192 slices (each is perpendicular to the line of sight). The small scale fluctuations consistent across  all models are due to sample variance. Indeed, since we average over a single thin slice, sample variance is much larger than when coeval snapshots are used. Moreover, since all models share the same initial conditions, and all lightcones used for this plot were generated using the same line of sight, ionized bubbles and other such features appear at the same location in each lightcone (hence the consistency the small scale fluctuations across the models).

\begin{figure*}
\begin{subfigure}{\textwidth}
\begin{picture}(100,60)
\put(0,0){\includegraphics[width=\textwidth]{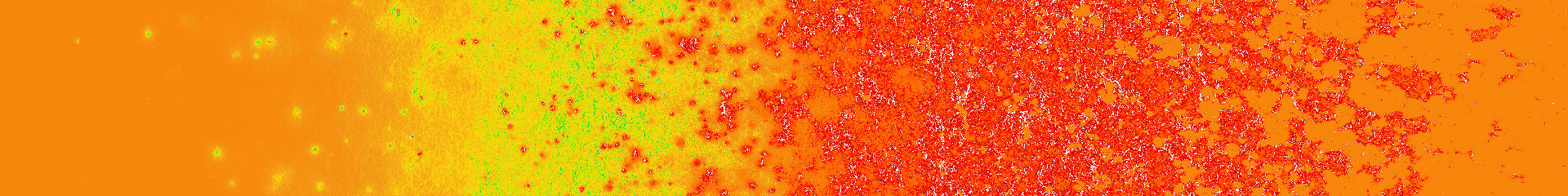}}
\put(5,51){\large $\bf f_X = 10$}
\put(5,38){\large $\bf r_{H/S} = 0$}
\put(5,22){\large $\bf f_\alpha = 2$}
\end{picture}
\end{subfigure}

\begin{subfigure}{\textwidth}
\includegraphics[width=\textwidth]{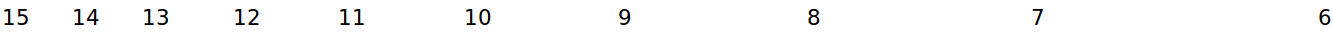}
\end{subfigure}

\begin{subfigure}{\textwidth}
\begin{picture}(100,64)
\put(0,0){\includegraphics[width=\textwidth]{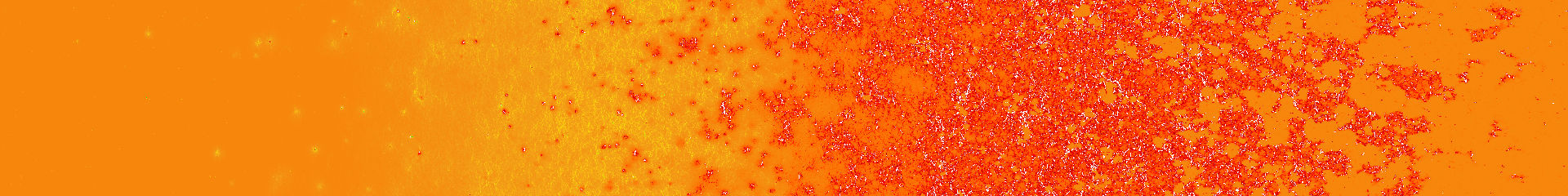}}
\put(5,51){\large $\bf f_X = 10$}
\put(5,38){\large $\bf r_{H/S} = 0$}
\put(5,22){\large $\bf f_\alpha = 0.5$}
\end{picture}
\end{subfigure}

\begin{subfigure}{\textwidth}
\includegraphics[width=\textwidth]{Redshift_Bar.png}
\end{subfigure}

\begin{subfigure}{\textwidth}
\begin{picture}(100,64)
\put(0,0){\includegraphics[width=\textwidth]{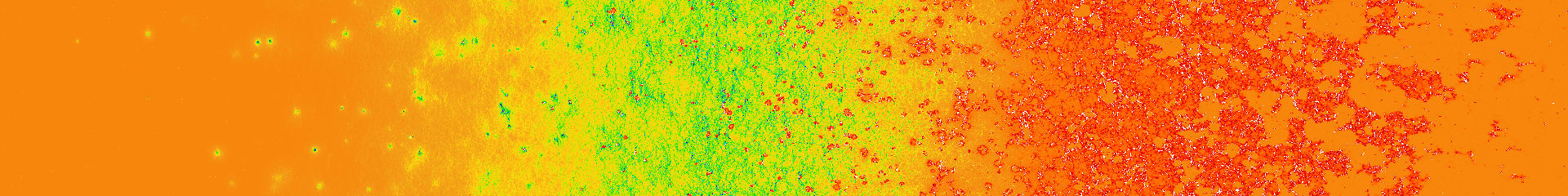}}
\put(5,51){\large $\bf f_X = 3$}
\put(5,38){\large $\bf r_{H/S} = 1$}
\put(5,22){\large $\bf f_\alpha = 1$}
\end{picture}
\end{subfigure}

\begin{subfigure}{\textwidth}
\includegraphics[width=\textwidth]{Redshift_Bar.png}
\end{subfigure}

\begin{subfigure}{\textwidth}
\begin{picture}(100,64)
\put(0,0){\includegraphics[width=\textwidth]{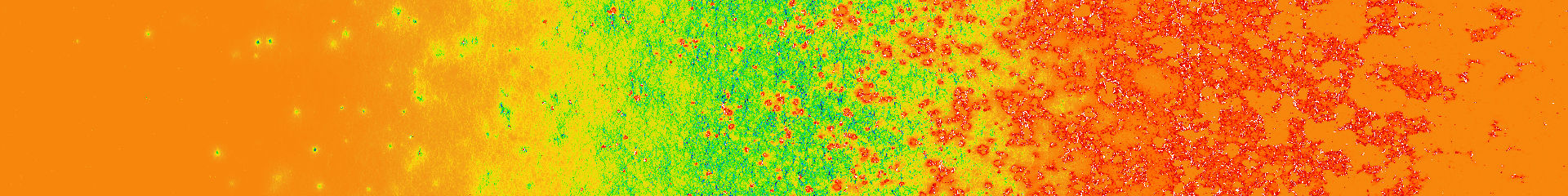}}
\put(5,51){\large $\bf f_X = 1$}
\put(5,38){\large $\bf r_{H/S} = 0$}
\put(5,22){\large $\bf f_\alpha = 1$}
\end{picture}
\end{subfigure}

\begin{subfigure}{\textwidth}
\includegraphics[width=\textwidth]{Redshift_Bar.png}
\end{subfigure}

\begin{subfigure}{\textwidth}
\begin{picture}(100,64)
\put(0,0){\includegraphics[width=\textwidth]{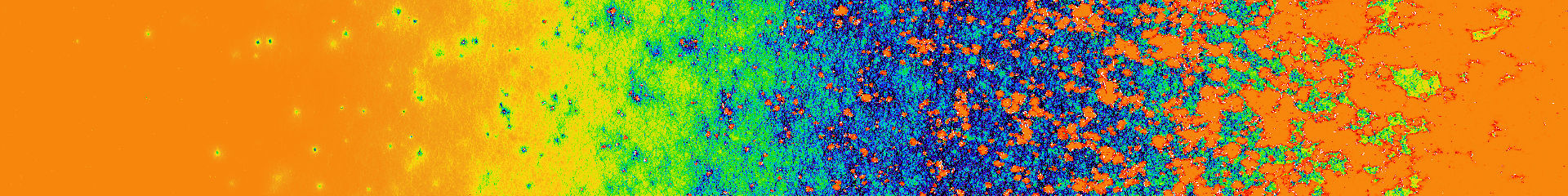}}
\put(5,51){\large $\bf f_X = 0.1$}
\put(5,38){\large $\bf r_{H/S} = 1$}
\put(5,22){\large $\bf f_\alpha = 1$}
\end{picture}
\end{subfigure}

\begin{subfigure}{\textwidth}
\includegraphics[width=\textwidth]{Redshift_Bar.png}
\end{subfigure}

\begin{subfigure}{\textwidth}
\begin{picture}(100,64)
\put(0,0){\includegraphics[width=\textwidth]{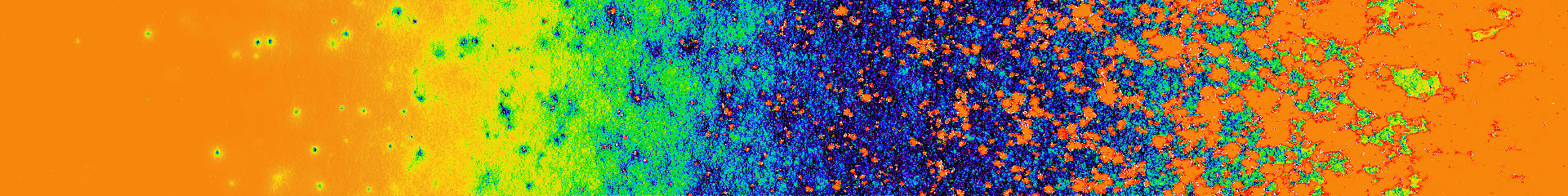}}
\put(5,51){\large $\bf f_X = 0.1$}
\put(5,38){\large $\bf r_{H/S} = 1$}
\put(5,22){\large $\bf f_\alpha = 2$}
\end{picture}
\end{subfigure}

\begin{subfigure}{\textwidth}
\includegraphics[width=\textwidth]{Redshift_Bar.png}
\end{subfigure}

\begin{subfigure}{\textwidth}
\includegraphics[width=\textwidth]{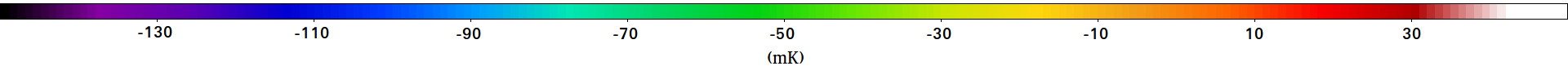}
\end{subfigure}
\caption{Brightness temperature maps for 6 models spanning the range of signals in the database (see labels on the maps) arranged from strongest emission (top) to strongest absorption (bottom). The maps are slices a single cell in thickness, taken parallel to the line of sight. They are made  using the high-resolution lightcones, and thus have a $1024\times8192$ resolution. Corresponding redshifts are indicated below the maps.}
\label{fig:lightcones}
\end{figure*}

Fig. \ref{mean_dtb} is mainly intended to give an impression of the  range of possible signals. However, for each value of $f_\alpha$, the specific model for each curve can be determined by looking at the strength of the absorption regime. Indeed, the latter is inversely proportional to the efficiency of the heating, and thus the vertical ordering is reversed from fig. \ref{mean_Tk} to fig. \ref{mean_dtb}. Let us emphasize again that we chose not to vary the source formation history in this first version of the database. It may be possible to  start source formation at higher redshift and rise more slowly while still satisfying the observational constraints. Then the absorption regime would be shifted towards higher $z$.

\subsection{Overview of the publicly available data}

\subsubsection{ Database web access}

The simulated data described below are available at \href{https://21ssd.obspm.fr/}{21ssd.obspm.fr}. They are available for download after a simple registration procedure (note that the largest files are $32$ GB). Instructions on how to read the files are provided as well as the source codes used to produce some of the data.
\subsubsection{High resolution 21-cm lightcones}
\label{sec:lightcones}
Lightcones were generated on-the-fly while running the simulations. They were made between the redshift of first source formation, $z=15$, and $z=6$. These lightcones are, in fact, a (large) set of particles with their associated redshift as well as all physical quantities necessary to compute the brightness temperature. The exception is the local Lyman-$\alpha$ flux lightcone. Since the corresponding simulation is run in post-processing and already interpolates between snapshots, the Lyman-$\alpha$ flux lightcone was also built by interpolation of snapshots, 10 Myr apart. The brightness temperature is then computed for each particle and mapped on a $1024 \times 1024 \times 8192$ grid (with the higher dimension along the line of sight) using SPH-like interpolation. The cells have equal $\delta a$ (or equivalently, $\delta \nu$) along the line of sight, and hence they vary in comoving thickness. The chosen resolution allows for near cubic cells in terms of comoving distance, with sizes close to the average inter-particle distance. A higher resolution would be useful in dense regions, but would produce files with a size unrealistic for standard internet downloads. As is, one lightcone is $32$ GB. The lower resolution lightcones, compatible with the expected SKA resolution, are computed by averaging over the high resolution ones. One high resolution lightcone is provided for each of the x, y and z observer directions and for each of the 45 models, for a total of 135 lightcones ($\sim 4$ TB of online data).

Fig. \ref{fig:lightcones} shows the same lightcone slices for a representative sample of 6 different models, ranging from those with a strong absorption regime to a very weak one. Of particular interest are the two middle panels that correspond to models with very similar average kinetic temperature histories, but somewhat different levels of brightness temperature fluctuations. 

Let us mention that the high resolution brightness temperature lightcones in the database contain a small number of cells with very strong emission ($> 100$ mK). These correspond to recombining high-density gas in galaxies, spread into elongated structures along the line of sight by redshift space distortions. While, in principle, this effect may actually be real and appear in observed data, it operates too close to our resolution limit for us to be confident in  its magnitude in our simulations. Database users may want to filter it out by clipping very high emission values. 

\subsubsection{SKA-resolution 21-cm lightcones including thermal noise}
\label{sec:SKA-resolution}

\begin{figure}
 \setlength{\fboxsep}{1pt}
  \centering
  { $ f_X = 10,\,\, r_{H/S} = 0,\,\, f_\alpha = 0.5$}
  \begin{minipage}[b]{0.49\textwidth}
  \begin{picture}(100,30)
	\put(0,0){\includegraphics[width=\textwidth]{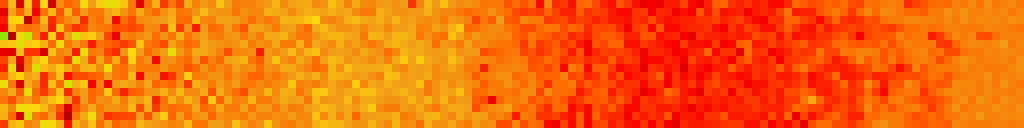}}
	   \put(2,21){ \fcolorbox{white}{white}{ $\!\! \Delta\theta = 6.1' \textrm{ - } 7.6'$}}
	\end{picture}
  \end{minipage}
  \hfill

  \vspace{-0.3cm}

  \begin{minipage}[b]{0.49\textwidth}
  \begin{picture}(100,30)
	\put(0,0){\includegraphics[width=\textwidth]{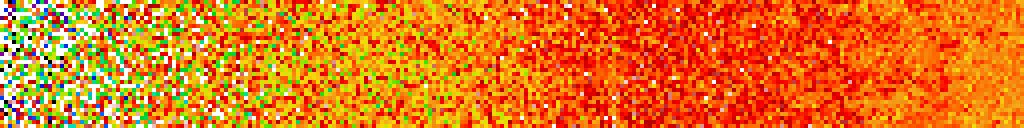}}
	 \put(2,21){ \fcolorbox{white}{white}{ $\!\! \Delta\theta = 3.1' \textrm{ - } 3.8'$}}
	\end{picture}
  \end{minipage}

  \vspace{0.1cm}

    { $ f_X = 1,\,\, r_{H/S} = 0,\,\, f_\alpha = 1$}
  \begin{minipage}[b]{0.49\textwidth}
  \begin{picture}(100,30)
	\put(0,0){\includegraphics[width=\textwidth]{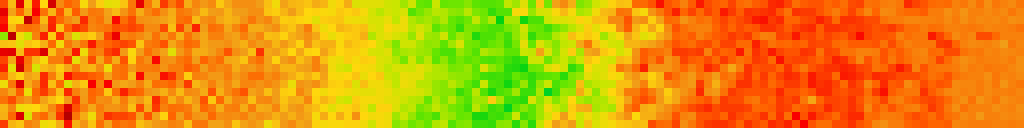}}
   \put(2,21){ \fcolorbox{white}{white}{ $\!\! \Delta\theta = 6.1' \textrm{ - } 7.6'$}}
	\end{picture}
  \end{minipage}
  \hfill

  \vspace{-0.3cm}

  \begin{minipage}[b]{0.49\textwidth}
  \begin{picture}(100,30)
	\put(0,0){\includegraphics[width=\textwidth]{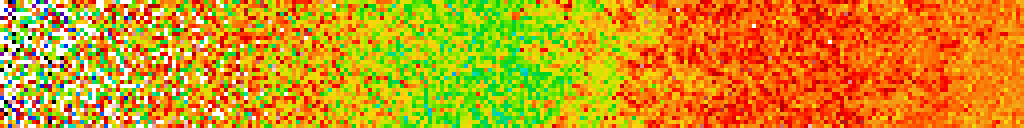}}
    \put(2,21){ \fcolorbox{white}{white}{ $\!\! \Delta\theta = 3.1' \textrm{ - } 3.8'$}}
	\end{picture}
  \end{minipage}

  \vspace{0.1cm}

    { $ f_X = 0.1,\,\, r_{H/S} = 1,\,\, f_\alpha = 2$}
  \begin{minipage}[b]{0.49\textwidth}
  \begin{picture}(100,30)
	\put(0,0){\includegraphics[width=\textwidth]{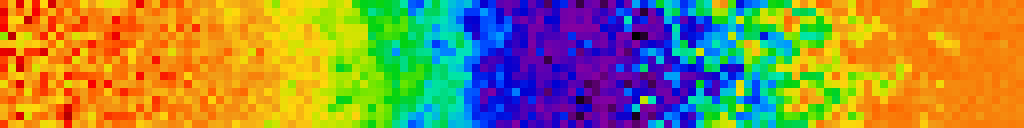}}
   \put(2,21){ \fcolorbox{white}{white}{ $\!\! \Delta\theta = 6.1' \textrm{ - } 7.6'$}}

	\end{picture}
  \end{minipage}
  \hfill

  \vspace{-0.3cm}

  \begin{minipage}[b]{0.49\textwidth}
  \begin{picture}(100,30)
	\put(0,0){\includegraphics[width=\textwidth]{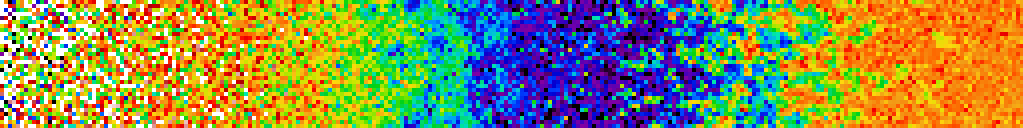}}
	\put(2,21){ \fcolorbox{white}{white}{ $\!\! \Delta\theta = 3.1' \textrm{ - } 3.8'$}}
	\end{picture}
  \end{minipage}
  
  \vspace{-0.1cm}

  \begin{minipage}[b]{0.49\textwidth}
    \includegraphics[width=\textwidth]{Redshift_Bar.png}
  \end{minipage}
  \hfill
  
  \vspace{-0.2cm}

\includegraphics[width=0.49\textwidth]{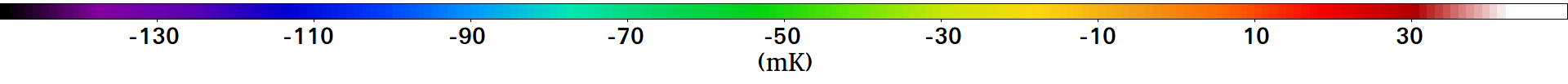}
\caption{Brightness temperature maps for 3 models identified by the labels on the maps, at $12.5$ and $6.25$ cMpc/h resolution (the corresponding angular resolution ranges are indicated on the maps). Thermal noise for a typical SKA survey (see main text for details) was added to the simulated signals. }
\label{fig:noisy_lightcones}
\end{figure}

From the high resolution lightcones, it is simple enough for future users of the database to produce lower resolution lightcones, to anticipate SKA tomographic capabilities. Regardless, for convenience we provide the corresponding sets of lightcones with resolutions of $32 \times 32 \times 256$ and $16 \times 16 \times 128$, corresponding to $6.25$ and $12.5$ cMpc/h or (varying) angular resolution on the order of $3^\prime$ and $6^\prime$, respectively. To produce these, we did not account for the beam shape. We simply computed the average of the high resolution cells within each low resolution one.  

We provide a data file containing the rms thermal noise level per pixel for the tomographic signal for a range of redshifts and angular resolutions. There are also realisations of the corresponding gaussian random variables in the form of thermal noise lightcones for the two low resolutions. The thermal noise level was computed using the SKA station configuration described in the SKA-TEL-SKO-0000422 document and thus it accounts for inhomogeneous UV coverage. The code for computing the noise is publicly available in the database, and survey parameters can be modified at will. The specific parameter used for the above realisations are the following: 1000h observing time (assuming 8h runs with the target field always at the meridian after 4h, which does not account for seasons), a field at $-30^\circ$ declination (compared to the $-26.8^\circ$ latitude of SKA-Low), and stations of diameter $35$ m, each containing $256$ dipoles with effective collecting area $\min(2.56,\lambda^{2 \over 3})$. We used $T_{\mathrm{sys}}=100 + 300 \left({\nu \over 150 \, \mathrm{MHz} } \right)^{-2.55}$ as in \citet{Mellema13}.

In fig. \ref{fig:noisy_lightcones} we show what the brightness temperature maps (sections of the lightcone parallel to the line of sight) look like after typical SKA noise is added. We consider the two resolutions mentioned above (the corresponding range of angular resolutions is shown on the top left of each map). It should be noted that the real SKA survey may produce maps with fixed angular resolution instead. In line with the estimation in \citet{Mellema13}, the $\sim 3^\prime$ resolution maps would not be usable in the case of a low intensity signal such as in the top panel. In all models tomography becomes difficult at $z>12$, unless a strong absorption regime occurs at these redshifts. The resolution required for a reasonable signal-to-noise ratio is typically $> 5^\prime$ ($10$ cMpc/h), which only allows us to map structures of (at least) several tens of cMpc in size. However, keep in mind that the actual SKA field is 3-dimensional and will cover a solid angle  about 10 times larger than ours. Thus, limitations at small scales will be compensated for by good statistics on large scales, and tomography 
should still prove a treasure trove of information.

\subsubsection{Power spectra}
\begin{figure*}
\includegraphics[width=0.48\textwidth]{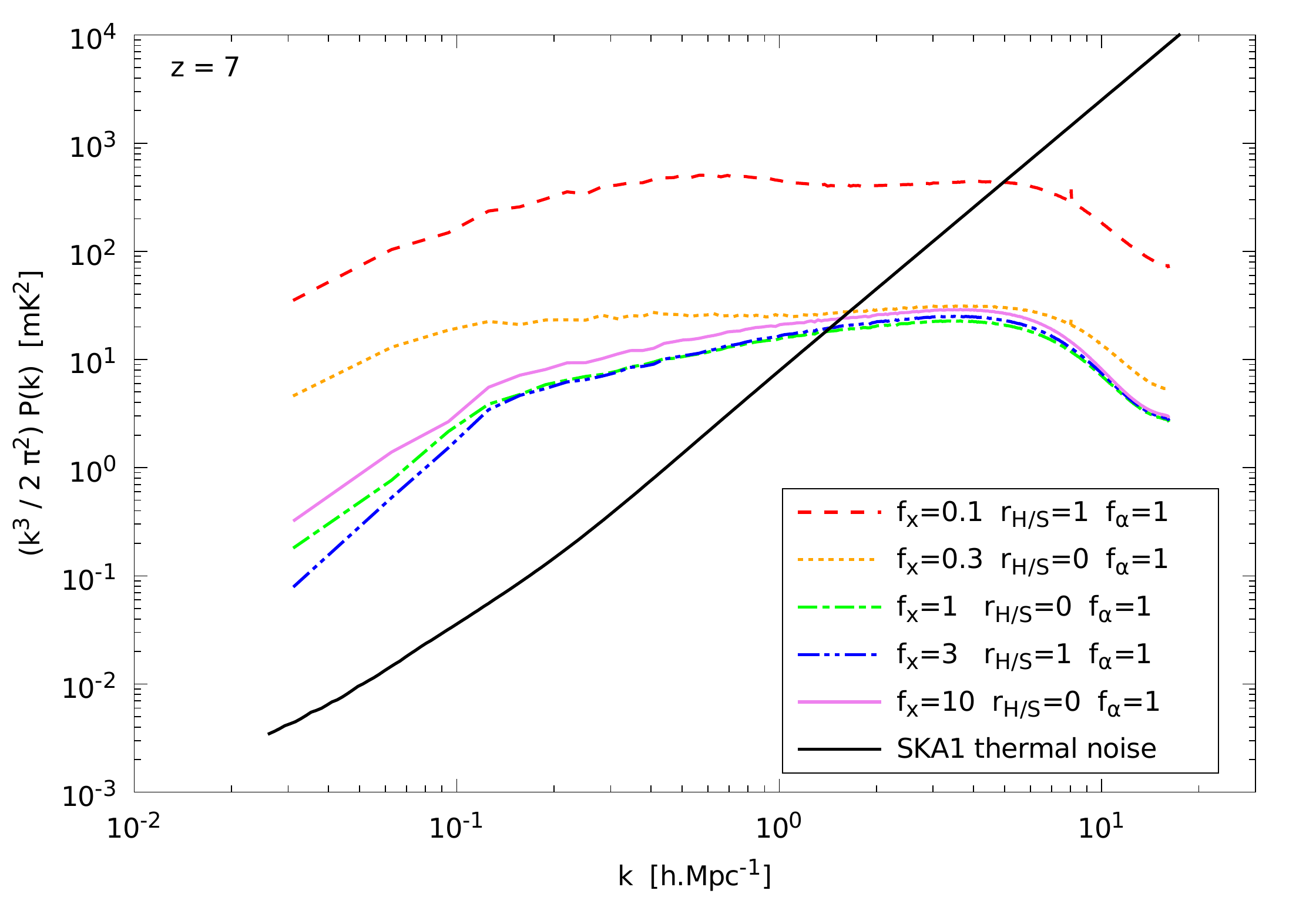}
\includegraphics[width=0.48\textwidth]{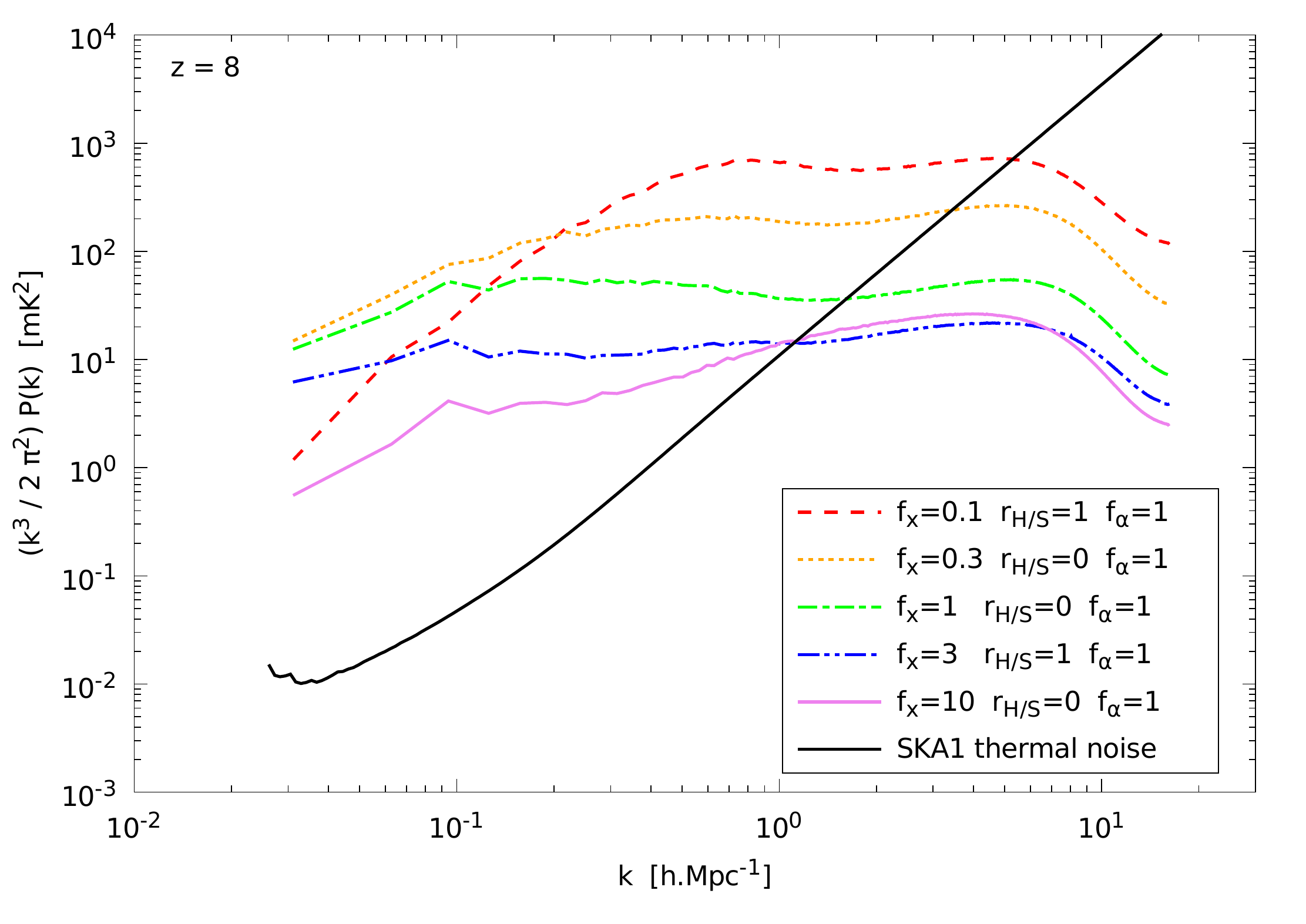}
\includegraphics[width=0.48\textwidth]{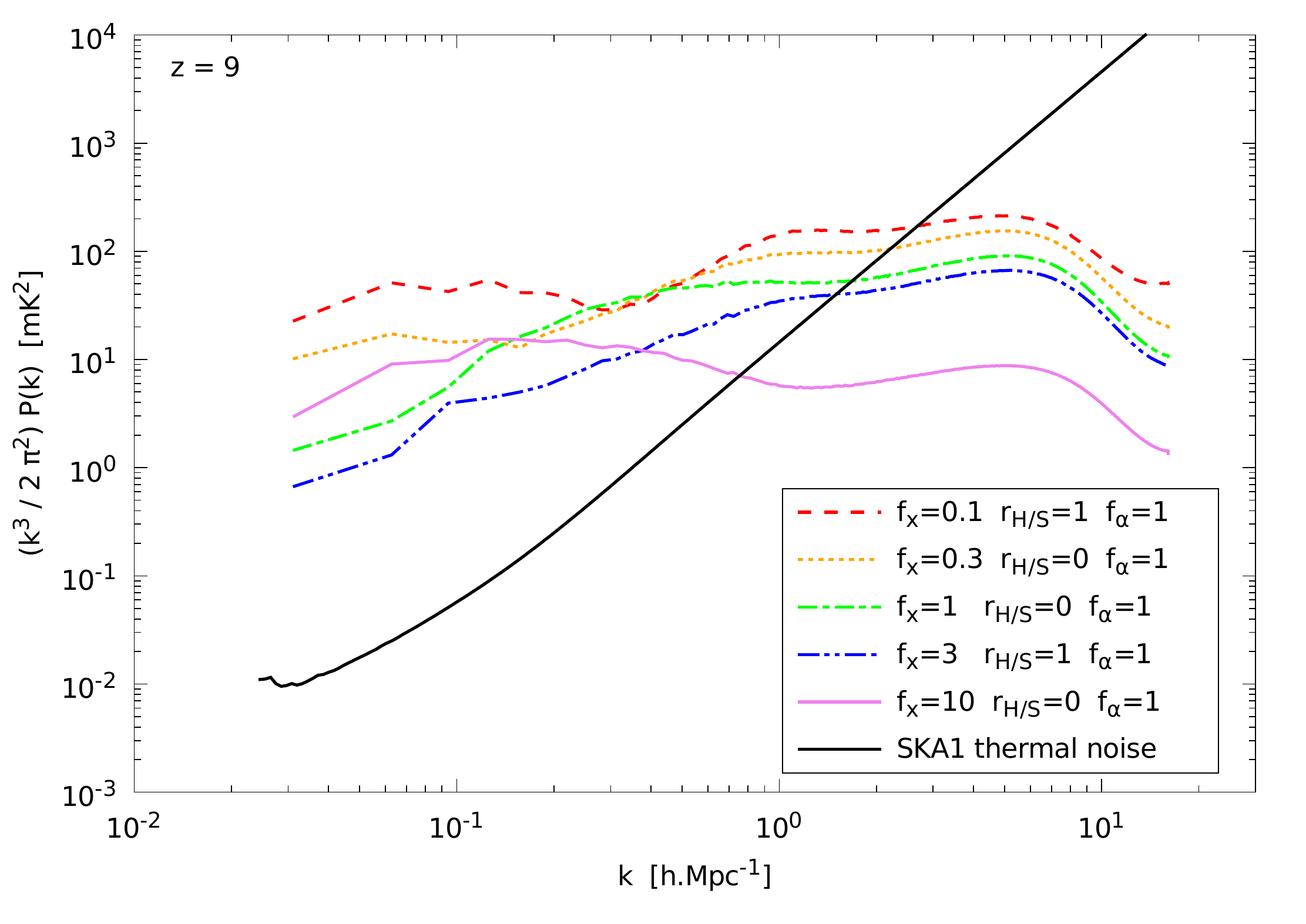}
\includegraphics[width=0.48\textwidth]{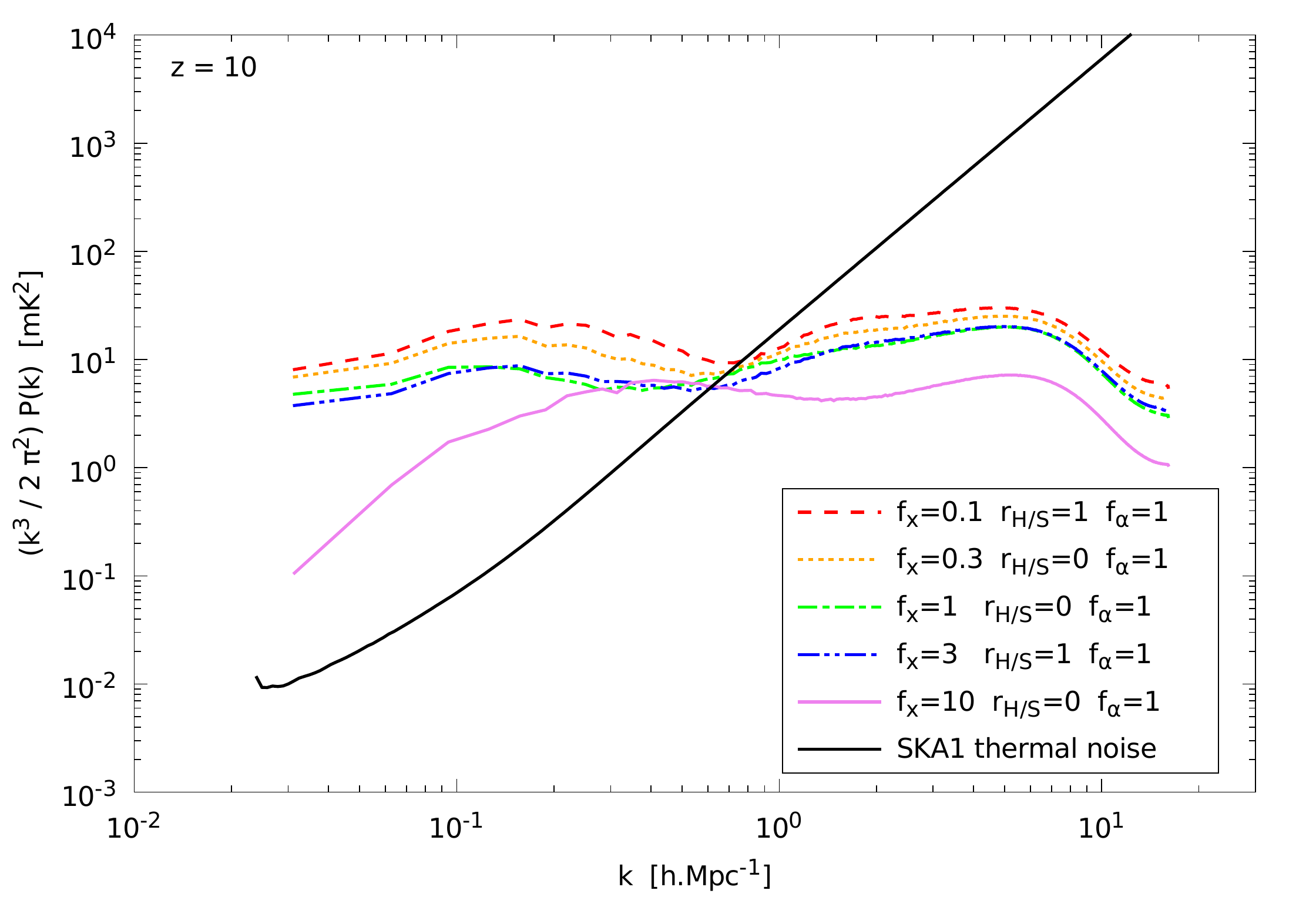}
\caption{Three-dimensional isotropic  power spectra at different redshifts (see labels) for 5 different models broadly covering the range of simulated signals. The thermal noise for a typical SKA survey is also plotted (see main text for survey parameters).}
\label{fig:powerspectrum}
\end{figure*}
We also provide 3D isotropic power spectra computed using the high-resolution lightcones (thus accounting for redshift space distortion and the evolution of the ionisation history). These are formatted as tabulated functions of $\|\bf{k}\|$ and $z$. The code used to compute the power spectra is available online. A sample of the variety of behaviour for different models in shown in fig. \ref{fig:powerspectrum}. It should be emphasized that the average amplitude of a given model at fixed $z$ depends strongly on the source formation history. A different history shifts  a given signal to a different $z$. Thus, relying on the amplitude to distinguish between models whose source formation history was keep constant is hardly possible: it is degenerate between different source formation histories. What may be more discriminating is the variation of the power spectrum as a function of the wavenumber: slope, fluctuations, etc. We can also see in fig. \ref{fig:powerspectrum} that,
while on large scales ($\|\bf{k}\| \sim 0.1$ h.cMpc$^{-1}$) the SKA sensitivity should allow us to easily distinguish between models, the small scale limit at which thermal noise overpowers the signal depends strongly on the model (and the source formation history). In our exploration of the parameter space, at $z=8$ this limit varies by a factor of more than $4$. The thermal noise in fig. \ref{fig:powerspectrum} was computed using the same survey parameters as in the previous section, a bandwidth of $10$ MHz, and a bin width of $\Delta k = k $. The code is also available on the database web site.

The provided power spectra and thermal noise levels can be used as templates. They can also be used as a training set to calibrate semi-numerical signal prediction methods, or to directly develop inversion algorithms requiring a sparse sampling of parameter space \citep[e.g.][]{Shimabukuro17}.

\subsubsection{Pixel distribution functions}
The power spectrum is a natural by-product  of interferometric observations and, being a statistical quantity, can be measured with a good signal to noise ratio. It does not, however, contain all the information when dealing with a non-Gaussian signal. On the other hand, tomography does. However, it also contains irrelevant information linked to the phases of the primordial random field (e.g. positions of the bubbles), and suffers from a high level of noise. Different statistical techniques have been suggested to quantify the non-Gaussian nature of the signal. The Pixel Distribution Function \citep{Ciardi03b,Mellema06b, Harker09, Ichikawa10, Baek10} is a diagnostic rich in information, especially in its redshift evolution. It could be used (but as of yet has not been) as an alternative to the power spectrum in parameter estimation methods \citep[e.g.][]{Shimabukuro17,Greig17,Kern17}. 

\begin{figure*}
\includegraphics[width=0.95\textwidth]{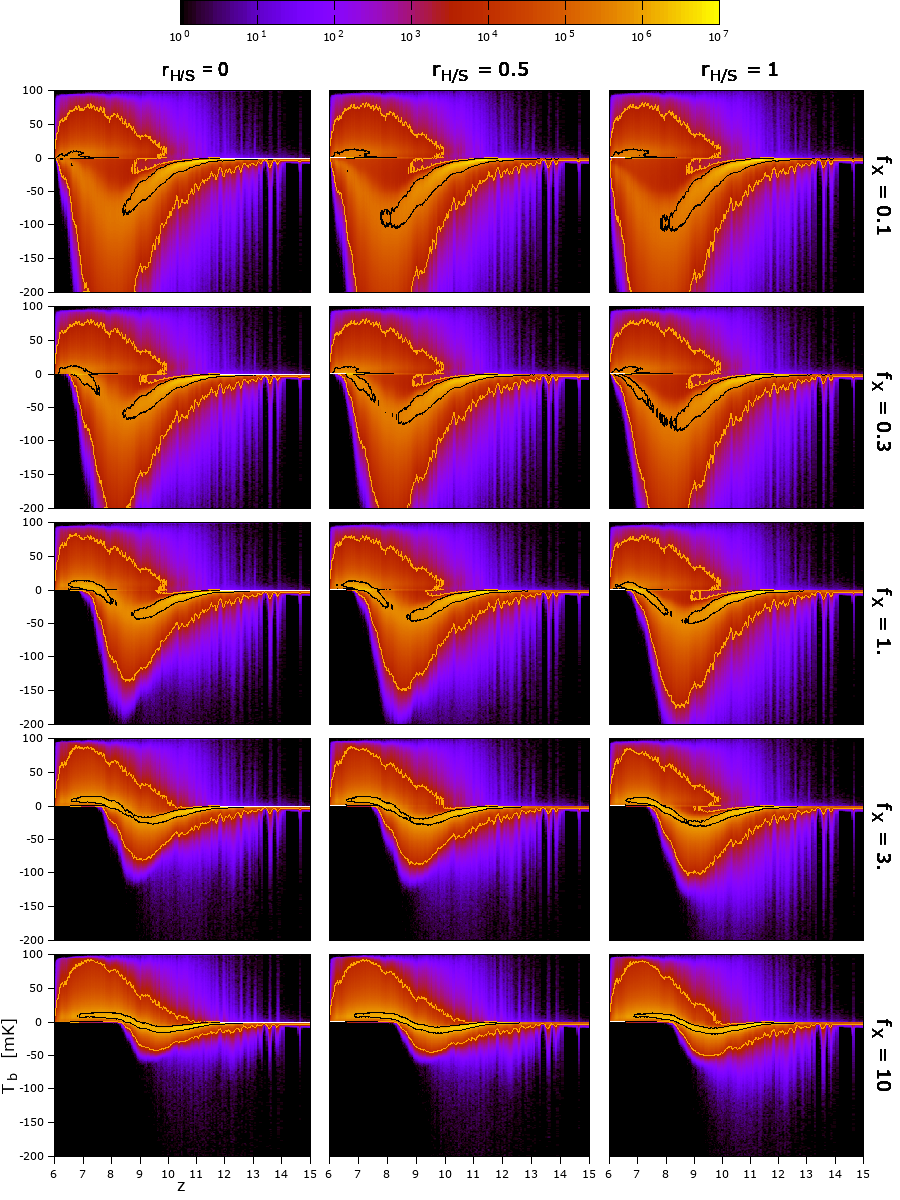}
\caption{Two dimensional histograms of the distribution of the 21-cm brightness temperature at different redshifts. This is made using the high resolution lightcones for all models with $f_\alpha=1$ (see labels for $f_X$ and $r_{H/S}$ values).  
This is a natural extension of the Pixel Distribution Function defined at a single redshift. The black and orange contours enclose regions where a pixel chosen at random in the lightcone will lie with 1$\sigma$ (0.682) and $3\sigma$ (0.997) confidence. Vertical spikes in the distribution are caused by sample variance. }
\label{fig:pdf}
\end{figure*}

In fig. \ref{fig:pdf} we present the two-dimensional PDFs (functions of brightness temperature and redshift) computed from the high-resolution lightcones for the $f_\alpha=1$ models in our database. We also plot 1-sigma and 3-sigma contour lines, meaning that a pixel chosen at random from the lightcones will lie within these contours with the corresponding confidence levels (we use fractions of pixels defined for a Gaussian distribution). We can see that both the PDFs and the contours depend strongly on the model, and hence they prove to be powerful discriminating tools, complementary to the power spectra (see section \ref{sec:distances}). 

However, the PDF is more sensitive to thermal noise than the power spectrum. For the power spectrum, long integration times and large survey volumes both help to reduce the thermal noise, and can be independently adjusted. This does not work for the PDF. By definition, each value of the power spectrum is an average of the Fourier mode amplitudes in a shell of fixed thickness. Increasing the Fourier space resolution (larger FoV) increases the number of cells in the shell, and the rms thermal noise on the average of the Fourier mode amplitudes decreases as the square root of the number of cells. In the case of the PDF, increasing the FoV decreases the sample variance, but leaves the thermal noise unchanged in each cell of the tomography. The effect of thermal noise on the PDF  is a vertical smearing, as seen in fig. \ref{fig:lowres_pdf}. In this figure we can see the  interdependent effects of thermal noise and limited resolution on the two extreme models `bracketing' our range of signals. The thermal noise is computed using the same SKA setup as in section \ref{sec:SKA-resolution}. As expected, we can see that when decreasing the resolution, we lose much of the faint purple plumes representing the less common brightness temperature values. Thus, the PDF becomes a less powerful diagnostic. Moreover, we can see that at $3^\prime$ resolution the SKA thermal noise will be of the same order or larger than the PDF dispersion  at all redshifts. At  $6^\prime$ resolution, the thermal noise remains smaller than the PDF dispersion at $z \lesssim 10$, and much smaller for the strong absorption model at $z \lesssim 8$. Let us finally mention that, when building the PDF from actual SKA tomographic data, the average brightness temperature at each redshift may not be available, removing part of the information contained in fig. \ref{fig:pdf} and \ref{fig:lowres_pdf}.

To summarise, while the theoretical PDF may hold as much information as the powerspectrum, perhaps even more, it is much more affected by instrumental limitations. We will need observations even deeper than a typically SKA survey to fully
uncover this information.

\setlength{\unitlength}{1.cm}
\begin{figure}
\begin{picture}(0.5\textwidth,0.692\textwidth)
\put(0,0){\includegraphics[width=0.48\textwidth]{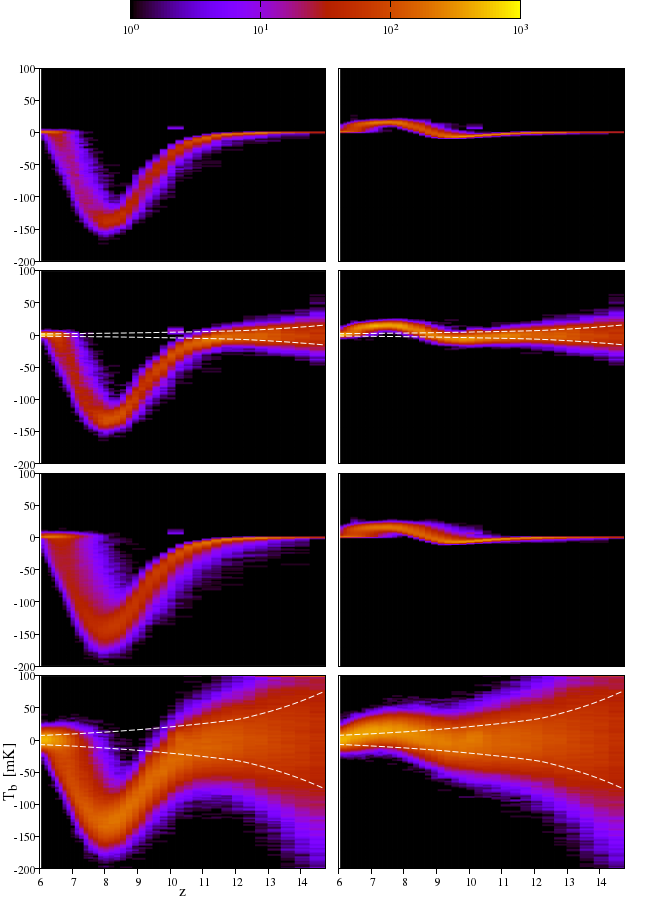}}
\put(0.05\textwidth,0.62\textwidth){ \scriptsize $f_X=0.1, \,r_{H/S} = 1, \,f_\alpha=2$}
\put(0.26\textwidth,0.62\textwidth){ \scriptsize $f_X=10, \,r_{H/S} = 1, \,f_\alpha=0.5$}
\put(0.345\textwidth,0.485\textwidth){ \small \bf \color{white} 6' resolution}
\put(0.275\textwidth,0.335\textwidth){ \small \bf \color{white} 6' resolution + noise}
\put(0.345\textwidth,0.185\textwidth){ \small \bf \color{white} 3' resolution }
\put(0.275\textwidth,0.035\textwidth){ \small \bf \color{white} 3' resolution + noise}
\put(0.125\textwidth,0.485\textwidth){ \small \bf \color{white} 6' resolution}
\put(0.055\textwidth,0.335\textwidth){ \small \bf \color{white} 6' resolution + noise}
\put(0.125\textwidth,0.185\textwidth){ \small \bf \color{white} 3' resolution }
\put(0.055\textwidth,0.035\textwidth){ \small \bf \color{white} 3' resolution + noise}

\end{picture}
\caption{Two dimensional PDF at two different SKA-like resolutions, with and without the corresponding thermal noise. The left and right columns are for two different models (see label on top): that with the strongest absorption and that with the strongest emission. The dashed line shows the level of thermal noise, which depends on resolution and redshift (see main text for survey details).}
\label{fig:lowres_pdf}
\end{figure}
\subsubsection{Data available on request}

While the spirit of 21SSD is to share data online, this puts a limit on the quantity of data that can be provided. The database contains less than 5 TB of data, but the raw output of the simulations is $\sim 100$ TB. For users who would like access to other types of data, and can arrange a reasonable transfer solution, we have the following quantities stored. First, we have coeval snapshots of the simulations every $10$, $20$ or $40$ Myr (depending on the model). These snapshots contain, for each particle, every quantity relevant to the physical processes included in the simulation  (position, velocity, density, temperature, ionisation fraction, etc). While the database offers gridded lightcones of the $21$ cm brightness temperature, the corresponding particle based lightcones are stored with the same quantities for each particle as with the snapshots. Finally, some runtime logs can be inspected to answer specific questions (e.g. the history of the average photon density). What is \textit{not} available are snapshots of the radiation field. LICORICE keeps the radiation field information for the last snapshot only (to allow a restart) in order to save disk space.

\section{Extracting knowledge from 21SSD: a first example}
\label{sec:distances}

\begin{figure}
\includegraphics[width=0.5\textwidth]{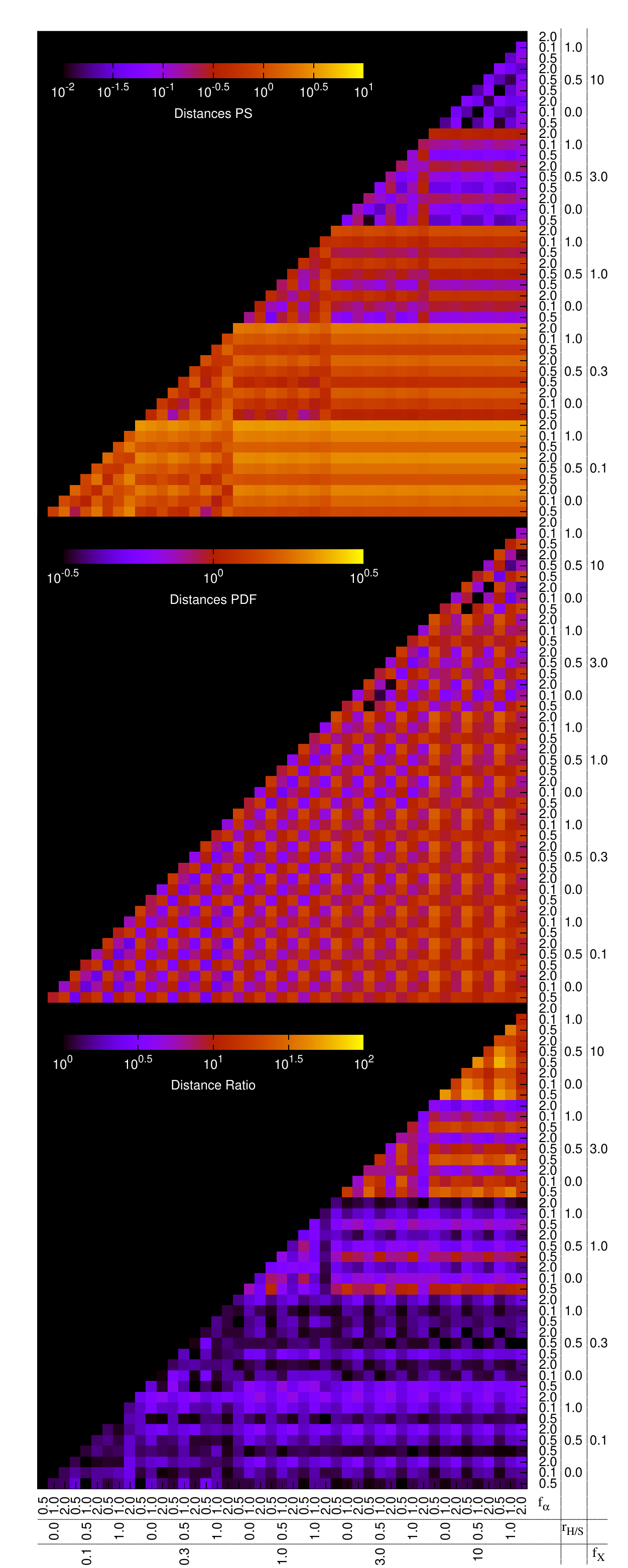}
\caption{Colour maps of the normalized power spectrum distance (top), normalized PDF distance (middle), and the ratio of these distances (bottom) for every pair of models in the database (see main text for the exact definitions). The $f_X$, $r_{H/S}$, and $f_\alpha$ values for each individual model can be found on the horizontal and vertical axes.}
\label{fig:Distances}
\end{figure}

The strength of the 21SSD database is the number of points explored in the chosen parameter space. Quantifying the variety of simulated signals, and assessing our capabilities to distinguish between them, is an important step towards preparing for the analysis of upcoming observations. Working with imaging data to distinguish
between models is difficult because they include random information (the specific positions and strengths of sources) that have to be filtered out. Working with the power spectrum or the PDF seems easier: we only need to beat sample variance (which is not fully achieved in our data, but will be with the SKA survey FoV). 

Using these diagnostics, it is necessary to first define a method of comparison between different models. The \emph{distance} between two simulations can be calculated using the $L_2$ norm (in which the power spectra or the pixel distribuation functions are treated as functions of two variables). Should the power spectra be used, the distance between models $i$ and $j$ is calculated as
\begin{equation}
\textrm{D}^{\textrm{PS}}_{i,j} = \left( \int{\left(\textrm{P}_i(k,z) - \textrm{P}_j(k,z)\right)^2dk dz} \right)^{1 \over 2} ~~~~~~
\end{equation}

\noindent where P$_i$ is the  power spectrum of the $i^{th}$ model, $k$ is the wave number in h.cMpc$^{-1}$, and $z$ is the redshift. In the case of the PDF, the calculation is similarly 
\begin{equation}
\textrm{D}^{\textrm{PDF}}_{i,j} = \left(\int\!\!{\left(\,\log(\textrm{D}_i(T_b,z)) - \log (\textrm{D}_j(T_b,z))\,\right)^2dT_b dz} \right)^{1 \over 2}
\end{equation}

\noindent where $\textrm{D}_i$ is the PDF of the $i^{th}$ model. The only difference in the definition is that we use the logarithm of the PDF. This is an arbitrary choice, however using the logarithm allows us to give more weight to the wings of the distributions, and more easily analyse the non-Gaussian features. We believe that this gives more discriminating power. More investigation is, however, needed to decide whether or not other definitions of distances would be preferable. Our goal here is to compare the discriminating power of the two diagnostics (power spectrum and PDF), not that of different distance definitions.

\subsection{Power spectra comparison}

The distance between the power spectra of all pairs of models in 21SSD is presented in the top panel of figure \ref{fig:Distances}, normalized by dividing by the average distance. As expected, this distance is largest between the model with parameters $[f_X,r_{H/S},f_\alpha]$ = [0.1, 1.0, 2.0] (strongest absorption) and simulations made with stronger  heating levels (different values of $f_X$). In general, it is seen that variations between each `block' of simulations made with equal $f_X$ are significant. However, within each of these blocks, the variations is weaker. This seems to suggest that using the power spectra to calculate distances discriminates well for $f_X$, but less powerfully for $r_{H/S}$ and $f_\alpha$, especially for high values of $f_X$.

\subsection{Pixel distribution functions comparison}

The PDF proves to be a very different diagnostic, which can be easily seen in the second panel of figure \ref{fig:Distances}. Firstly, comparing the PDFs leads to a smaller range of values for distance. This is because each individual PDF has the same $L_1$ norm (regardless of the model there are a constant number of pixels in each lightcone). Therefore, the distance between two PDFs is somewhat more constrained. Conversely, the power spectra can be orders of magnitude lower or higher than others, depending on the efficiency of heating. This difference explains the fact that the normalized power spectra distances fall between three orders of magnitude ($10^{-2}$ - $10^{1}$) while the normalized PDF distance only cover roughly one order ($10^{-0.5}$ - $10^{0.5}$).

In addition to magnitude, the PDF distances are also very different in form. There is little variation from one 9x9 `block' of constant $f_X$ values to another, nor between 3x3 blocks of constant $r_{H/S}$, yet there is substantial difference between simulations made with different $f_\alpha$ values. This is seen in the different magnitudes of neighbouring pixels.

\subsection{Comparing methods}
To properly examine where the different diagnostics excel,  we define the \emph{Distance Ratio} as:
\begin{equation}
\textrm{Ratio}(i,j) = \max\left(\frac{\widetilde{\textrm{D}}^{\textrm{PS}}_{i,j}}{\widetilde{\textrm{D}}^{\textrm{PDF}}_{i,j}}, \frac{\widetilde{\textrm{D}}^{\textrm{PDF}}_{i,j}}{\widetilde{\textrm{D}}^{\textrm{PS}}_{i,j}} \right)
\end{equation}
where $\widetilde{\textrm{D}}$ are normalized distances. A high value for this ratio means that the two diagnostics have very different discriminating power.
We can see the map of this quantity in the third panel of figure \ref{fig:Distances}. The main information in this map is that the difference in  discriminating power is especially strong between models with high heating levels (upper part of the triangle). One may be tempted to conclude that the PDF is a better tool  in this region of parameter space, however this would be premature. The two diagnostics have different sensitivity to the noise, and even in regions where the PDF appears to perform better, this advantage may be lost to the noise the PDF experiences over the power spectrum. A final conclusion would need to account for the noise. That, along with exploring different distance definitions, will be addressed in a future work. Let us finally note that the structure in $f_\alpha$ is also seen to stand out, especially at $f_X$ = 0.1, thanks to the sensitivity of the PDF to this parameter.

These results provide a first idea of how best to utilize the power spectra and PDF distances to extract parameters. If thermal noise is neglected  the power spectrum distance is a better tracer for the $f_X$ parameter, while the PDF distance is more suited to determine the $f_\alpha$ value. Neither the power spectrum nor the PDF seem especially powerful tracers for the $r_{H/S}$ variable -- perhaps somewhat for the power spectrum distances, where slightly different distances are seen between simulations varying only in $r_{H/S}$. It is possible that a robust parameter extraction system will have to take advantage of both the power spectra and the PDF.

Some very tentative hints towards optimal parameter space sampling are present. The similarity between the $f_X$ = 10 and 3 regions of the Distance Ratios (as well as, of course, the power spectrum and PDF distance figures) indicates that a more sparse sampling in this region would have been possible. Using this full distance information to derive an optimal parameter space sampling will be explored further in a subsequent article.

\section{Conclusions}

In this work we have presented a publicly available database of simulated 21-cm EoR signals. High-resolution radiative hydrodynamic simulations ($1024^3$ particles in a $200$ Mpc.h$^{-1}$) were performed with the LICORICE code to predict the signal. The modelling includes radiative transfer in the ionizing UV and X-ray bands coupled to the dynamics, as well as in the Lyman lines (to accurately account for the Wouthuysen-Field effect) in post-processing. With the ionization history unchanged (fitting the available observational constraints) a three-dimensional parameter space is explored wich contains 45 different models. This is accomplished by varying the X-ray production efficiency (the $f_X$ parameter), the X-ray spectral properties (the $r_{H/S}$ parameter), and the Lyman band photon production efficiency (the $f_\alpha$ parameter). Thus, we mainly explore the processes responsible for fluctuations of the $21$-cm signal due to heating and Lyman-$\alpha$ coupling. In a future extension of the database it may be relevant to explore different average ionization histories and ionization fluctuations (within the available observational constraints).

We presented the data products available in the database, which currently include high resolution and typical SKA-like resolution lightcones in three directions for each of the $45$ models, their associated power spectra and two-dimensional Pixel Distribution Functions, as well as thermal noise models for the SKA. The source code used to derive these quantities is also available. We have given some examples of power spectra, compared them to typical SKA noise, and studied the two-dimensional PDF in some details, investigated mainly in one dimension in previous studies. We showed in particular that, while the theoretical (noise-free) PDF potentially contains a wealth of information not seen in the power spectrum, taking SKA-like thermal noise and resolution into account substantially decreases the amount of usable information. 

Finally, we took a first attempt at quantifying the degree to which different diagnostics respond to model variation. This study is preliminary and needs to also take into account instrumental limitations (primarily resolution and noise). 

\section*{Acknowledgments}
This work was made in the framework of the French ANR funded project ORAGE (ANR-14-CE33-0016). We also acknowledge the support of the ILP LABEX (under the reference ANR-10-LABX-63) within the Investissements
d'Avenir programme under reference ANR-11-IDEX-0004-02. The simulations were performed on the GENCI
national computing center at CCRT and CINES (DARI grants number 2014046667 and 2015047376).

We would also like to thank Keith Grainge, whose wisdom set us on the right track when preparing to realistically simulate SKA-level noise.
 \bibliographystyle{mnras}
\bibliography{myref}

\begin{thebibliography}{}
\makeatletter
\relax
\def\mn@urlcharsother{\let\do\@makeother \do\$\do\&\do\#\do\^\do\_\do\%\do\~}
\def\mn@doi{\begingroup\mn@urlcharsother \@ifnextchar [ {\mn@doi@}
  {\mn@doi@[]}}
\def\mn@doi@[#1]#2{\def\@tempa{#1}\ifx\@tempa\@empty \href
  {http://dx.doi.org/#2} {doi:#2}\else \href {http://dx.doi.org/#2} {#1}\fi
  \endgroup}
\def\mn@eprint#1#2{\mn@eprint@#1:#2::\@nil}
\def\mn@eprint@arXiv#1{\href {http://arxiv.org/abs/#1} {{\tt arXiv:#1}}}
\def\mn@eprint@dblp#1{\href {http://dblp.uni-trier.de/rec/bibtex/#1.xml}
  {dblp:#1}}
\def\mn@eprint@#1:#2:#3:#4\@nil{\def\@tempa {#1}\def\@tempb {#2}\def\@tempc
  {#3}\ifx \@tempc \@empty \let \@tempc \@tempb \let \@tempb \@tempa \fi \ifx
  \@tempb \@empty \def\@tempb {arXiv}\fi \@ifundefined
  {mn@eprint@\@tempb}{\@tempb:\@tempc}{\expandafter \expandafter \csname
  mn@eprint@\@tempb\endcsname \expandafter{\@tempc}}}

\bibitem[\protect\citeauthoryear{{Ali} et~al.,}{{Ali} et~al.}{2015}]{Ali15}
{Ali} Z.~S.,  et~al., 2015, \mn@doi [ApJ] {10.1088/0004-637X/809/1/61}, \href
  {http://adsabs.harvard.edu/abs/2015ApJ...809...61A} {809, 61}

\bibitem[\protect\citeauthoryear{Baek, Di~Matteo, Semelin, Combes  \&
  Revaz}{Baek et~al.}{2009}]{Baek09}
Baek S.,  Di~Matteo P.,  Semelin B.,  Combes F.,   Revaz Y.,  2009, A\&A, 495,
  389

\bibitem[\protect\citeauthoryear{{Baek}, {Semelin}, {Di Matteo}, {Revaz}  \&
  {Combes}}{{Baek} et~al.}{2010}]{Baek10}
{Baek} S.,  {Semelin} B.,  {Di Matteo} P.,  {Revaz} Y.,   {Combes} F.,  2010,
  \mn@doi [A\&A] {10.1051/0004-6361/201014347}, \href
  {http://adsabs.harvard.edu/abs/2010A%26A...523A...4B} {523, A4+}

\bibitem[\protect\citeauthoryear{{Bouwens} et~al.,}{{Bouwens}
  et~al.}{2015a}]{Bouwens15b}
{Bouwens} R.~J.,  et~al., 2015a, \mn@doi [\apj] {10.1088/0004-637X/803/1/34},
  \href {http://adsabs.harvard.edu/abs/2015ApJ...803...34B} {803, 34}

\bibitem[\protect\citeauthoryear{{Bouwens}, {Illingworth}, {Oesch}, {Caruana},
  {Holwerda}, {Smit}  \& {Wilkins}}{{Bouwens} et~al.}{2015b}]{Bouwens15a}
{Bouwens} R.~J.,  {Illingworth} G.~D.,  {Oesch} P.~A.,  {Caruana} J.,
  {Holwerda} B.,  {Smit} R.,   {Wilkins} S.,  2015b, \mn@doi [\apj]
  {10.1088/0004-637X/811/2/140}, \href
  {http://adsabs.harvard.edu/abs/2015ApJ...811..140B} {811, 140}

\bibitem[\protect\citeauthoryear{Ciardi \& Madau}{Ciardi \&
  Madau}{2003}]{Ciardi03b}
Ciardi B.,  Madau P.,  2003, ApJ, 596, 1

\bibitem[\protect\citeauthoryear{{Cohen}, {Fialkov}, {Barkana}  \&
  {Lotem}}{{Cohen} et~al.}{2016}]{Cohen16}
{Cohen} A.,  {Fialkov} A.,  {Barkana} R.,   {Lotem} M.,  2016, preprint, \href
  {http://adsabs.harvard.edu/abs/2016arXiv160902312C} {} (\mn@eprint {arXiv}
  {1609.02312})

\bibitem[\protect\citeauthoryear{{Dillon} et~al.,}{{Dillon}
  et~al.}{2015}]{Dillon15}
{Dillon} J.~S.,  et~al., 2015, \mn@doi [PhRevD] {10.1103/PhysRevD.91.123011},
  \href {http://adsabs.harvard.edu/abs/2015PhRvD..91l3011D} {91, 123011}

\bibitem[\protect\citeauthoryear{{Ewall-Wice}, {Hewitt}, {Mesinger}, {Dillon},
  {Liu}  \& {Pober}}{{Ewall-Wice} et~al.}{2016a}]{Ewall-Wice16b}
{Ewall-Wice} A.,  {Hewitt} J.,  {Mesinger} A.,  {Dillon} J.~S.,  {Liu} A.,
  {Pober} J.,  2016a, \mn@doi [MNRAS] {10.1093/mnras/stw452}, \href
  {http://adsabs.harvard.edu/abs/2016MNRAS.458.2710E} {458, 2710}

\bibitem[\protect\citeauthoryear{{Ewall-Wice} et~al.,}{{Ewall-Wice}
  et~al.}{2016b}]{Ewall-Wice16}
{Ewall-Wice} A.,  et~al., 2016b, \mn@doi [MNRAS] {10.1093/mnras/stw1022}, \href
  {http://adsabs.harvard.edu/abs/2016MNRAS.460.4320E} {460, 4320}

\bibitem[\protect\citeauthoryear{{Fan} et~al.,}{{Fan} et~al.}{2006}]{Fan06}
{Fan} X.,  et~al., 2006, \mn@doi [AJ] {10.1086/504836}, \href
  {http://cdsads.u-strasbg.fr/abs/2006AJ....132..117F} {132, 117}

\bibitem[\protect\citeauthoryear{{Fialkov}, {Barkana}  \& {Visbal}}{{Fialkov}
  et~al.}{2014}]{Fialkov14}
{Fialkov} A.,  {Barkana} R.,   {Visbal} E.,  2014, \mn@doi [Nature]
  {10.1038/nature12999}, \href
  {http://cdsads.u-strasbg.fr/abs/2014Natur.506..197F} {506, 197}

\bibitem[\protect\citeauthoryear{Field}{Field}{1958}]{Field58}
Field G.,  1958, Proc. IRE, 46, 240

\bibitem[\protect\citeauthoryear{{Fragos} et~al.,}{{Fragos}
  et~al.}{2013}]{Fragos13}
{Fragos} T.,  et~al., 2013, \mn@doi [ApJ] {10.1088/0004-637X/764/1/41}, \href
  {http://cdsads.u-strasbg.fr/abs/2013ApJ...764...41F} {764, 41}

\bibitem[\protect\citeauthoryear{{Furlanetto}}{{Furlanetto}}{2006}]{Furlanetto06d}
{Furlanetto} S.~R.,  2006, \mn@doi [MNRAS] {10.1111/j.1365-2966.2006.10725.x},
  \href {http://adsabs.harvard.edu/abs/2006MNRAS.371..867F} {371, 867}

\bibitem[\protect\citeauthoryear{Furlanetto, Oh  \& Briggs}{Furlanetto
  et~al.}{2006}]{Furlanetto06}
Furlanetto S.~R.,  Oh S.~P.,   Briggs F.~H.,  2006, PhR, 433, 181

\bibitem[\protect\citeauthoryear{{Greig} \& {Mesinger}}{{Greig} \&
  {Mesinger}}{2015}]{Greig15}
{Greig} B.,  {Mesinger} A.,  2015, \mn@doi [MNRAS] {10.1093/mnras/stv571},
  \href {http://adsabs.harvard.edu/abs/2015MNRAS.449.4246G} {449, 4246}

\bibitem[\protect\citeauthoryear{{Greig} \& {Mesinger}}{{Greig} \&
  {Mesinger}}{2017}]{Greig17}
{Greig} B.,  {Mesinger} A.,  2017, preprint, \href
  {http://adsabs.harvard.edu/abs/2017arXiv170503471G} {} (\mn@eprint {arXiv}
  {1705.03471})

\bibitem[\protect\citeauthoryear{{Hahn} \& {Abel}}{{Hahn} \&
  {Abel}}{2011}]{Hahn11}
{Hahn} O.,  {Abel} T.,  2011, \mn@doi [MNRAS]
  {10.1111/j.1365-2966.2011.18820.x}, \href
  {http://adsabs.harvard.edu/abs/2011MNRAS.415.2101H} {415, 2101}

\bibitem[\protect\citeauthoryear{{Harker} et~al.,}{{Harker}
  et~al.}{2009}]{Harker09}
{Harker} G.,  et~al., 2009, \mn@doi [MNRAS] {10.1111/j.1365-2966.2009.15081.x},
  \href {http://cdsads.u-strasbg.fr/abs/2009MNRAS.397.1138H} {397, 1138}

\bibitem[\protect\citeauthoryear{Hirata}{Hirata}{2006}]{Hirata06}
Hirata C.~M.,  2006, MNRAS, 367, 259

\bibitem[\protect\citeauthoryear{Ichikawa, Barkana, Iliev, Mellema  \&
  Shapiro}{Ichikawa et~al.}{2010}]{Ichikawa10}
Ichikawa K.,  Barkana R.,  Iliev I.~T.,  Mellema G.,   Shapiro P.,  2010,
  MNRAS, 406, 2521

\bibitem[\protect\citeauthoryear{Iliev, Whalen, Mellema, Ahn  \& Baek}{Iliev
  et~al.}{2009}]{Iliev09}
Iliev I.~T.,  Whalen D.,  Mellema G.,  Ahn K.,   Baek S.,  2009, MNRAS, 400,
  1283

\bibitem[\protect\citeauthoryear{{Iliev}, {Mellema}, {Ahn}, {Shapiro}, {Mao}
  \& {Pen}}{{Iliev} et~al.}{2014}]{Iliev14}
{Iliev} I.~T.,  {Mellema} G.,  {Ahn} K.,  {Shapiro} P.~R.,  {Mao} Y.,   {Pen}
  U.-L.,  2014, \mn@doi [MNRAS] {10.1093/mnras/stt2497}, \href
  {http://adsabs.harvard.edu/abs/2014MNRAS.439..725I} {439, 725}

\bibitem[\protect\citeauthoryear{{Kern}, {Liu}, {Parsons}, {Mesinger}  \&
  {Greig}}{{Kern} et~al.}{2017}]{Kern17}
{Kern} N.~S.,  {Liu} A.,  {Parsons} A.~R.,  {Mesinger} A.,   {Greig} B.,  2017,
  preprint, \href {http://adsabs.harvard.edu/abs/2017arXiv170504688K} {}
  (\mn@eprint {arXiv} {1705.04688})

\bibitem[\protect\citeauthoryear{{Liu} \& {Parsons}}{{Liu} \&
  {Parsons}}{2016}]{Liu16}
{Liu} A.,  {Parsons} A.~R.,  2016, \mn@doi [MNRAS] {10.1093/mnras/stw071},
  \href {http://adsabs.harvard.edu/abs/2016MNRAS.457.1864L} {457, 1864}

\bibitem[\protect\citeauthoryear{{McGreer}, {Mesinger}  \&
  {D'Odorico}}{{McGreer} et~al.}{2015}]{McGreer15}
{McGreer} I.~D.,  {Mesinger} A.,   {D'Odorico} V.,  2015, \mn@doi [MNRAS]
  {10.1093/mnras/stu2449}, \href
  {http://adsabs.harvard.edu/abs/2015MNRAS.447..499M} {447, 499}

\bibitem[\protect\citeauthoryear{Mellema, Iliev, Pen  \& Shapiro}{Mellema
  et~al.}{2006}]{Mellema06b}
Mellema G.,  Iliev I.~T.,  Pen U.-L.,   Shapiro P.~R.,  2006, MNRAS, 372, 679

\bibitem[\protect\citeauthoryear{{Mellema} et~al.,}{{Mellema}
  et~al.}{2013}]{Mellema13}
{Mellema} G.,  et~al., 2013, \mn@doi [ExA] {10.1007/s10686-013-9334-5}, \href
  {http://cdsads.u-strasbg.fr/abs/2013ExA...tmp...17M} {}

\bibitem[\protect\citeauthoryear{{Mesinger}, {Greig}  \& {Sobacchi}}{{Mesinger}
  et~al.}{2016}]{Mesinger16}
{Mesinger} A.,  {Greig} B.,   {Sobacchi} E.,  2016, \mn@doi [MNRAS]
  {10.1093/mnras/stw831}, \href
  {http://adsabs.harvard.edu/abs/2016MNRAS.459.2342M} {459, 2342}

\bibitem[\protect\citeauthoryear{{Ocvirk} et~al.,}{{Ocvirk}
  et~al.}{2016}]{Ocvirk16}
{Ocvirk} P.,  et~al., 2016, \mn@doi [MNRAS] {10.1093/mnras/stw2036}, \href
  {http://adsabs.harvard.edu/abs/2016MNRAS.463.1462O} {463, 1462}

\bibitem[\protect\citeauthoryear{{Paciga} et~al.,}{{Paciga}
  et~al.}{2013}]{Paciga13}
{Paciga} G.,  et~al., 2013, preprint, \href
  {http://cdsads.u-strasbg.fr/abs/2013arXiv1301.5906P} {} (\mn@eprint {arXiv}
  {1301.5906})

\bibitem[\protect\citeauthoryear{{Patil} et~al.,}{{Patil}
  et~al.}{2017}]{Patil17}
{Patil} A.~H.,  et~al., 2017, \mn@doi [ApJ] {10.3847/1538-4357/aa63e7}, \href
  {http://adsabs.harvard.edu/abs/2017ApJ...838...65P} {838, 65}

\bibitem[\protect\citeauthoryear{{Planck Collaboration} et~al.,}{{Planck
  Collaboration} et~al.}{2015}]{Planck15}
{Planck Collaboration} et~al., 2015, preprint, \href
  {http://adsabs.harvard.edu/abs/2015arXiv150201589P} {} (\mn@eprint {arXiv}
  {1502.01589})

\bibitem[\protect\citeauthoryear{{Pober} et~al.,}{{Pober}
  et~al.}{2014}]{Pober14}
{Pober} J.~C.,  et~al., 2014, \mn@doi [ApJ] {10.1088/0004-637X/782/2/66}, \href
  {http://adsabs.harvard.edu/abs/2014ApJ...782...66P} {782, 66}

\bibitem[\protect\citeauthoryear{{Pober} et~al.,}{{Pober}
  et~al.}{2015}]{Pober15}
{Pober} J.~C.,  et~al., 2015, \mn@doi [ApJ] {10.1088/0004-637X/809/1/62}, \href
  {http://adsabs.harvard.edu/abs/2015ApJ...809...62P} {809, 62}

\bibitem[\protect\citeauthoryear{{Schenker}, {Ellis}, {Konidaris}  \&
  {Stark}}{{Schenker} et~al.}{2014}]{Schenker14}
{Schenker} M.~A.,  {Ellis} R.~S.,  {Konidaris} N.~P.,   {Stark} D.~P.,  2014,
  \mn@doi [ApJ] {10.1088/0004-637X/795/1/20}, \href
  {http://adsabs.harvard.edu/abs/2014ApJ...795...20S} {795, 20}

\bibitem[\protect\citeauthoryear{{Schroeder}, {Mesinger}  \&
  {Haiman}}{{Schroeder} et~al.}{2013}]{Schroeder13}
{Schroeder} J.,  {Mesinger} A.,   {Haiman} Z.,  2013, \mn@doi [MNRAS]
  {10.1093/mnras/sts253}, \href
  {http://adsabs.harvard.edu/abs/2013MNRAS.428.3058S} {428, 3058}

\bibitem[\protect\citeauthoryear{{Semelin}}{{Semelin}}{2016}]{Semelin16}
{Semelin} B.,  2016, \mn@doi [MNRAS] {10.1093/mnras/stv2312}, \href
  {http://adsabs.harvard.edu/abs/2016MNRAS.455..962S} {455, 962}

\bibitem[\protect\citeauthoryear{Semelin \& Combes}{Semelin \&
  Combes}{2002}]{Semelin02}
Semelin B.,  Combes F.,  2002, A\&A, 495, 389

\bibitem[\protect\citeauthoryear{Semelin, Combes  \& Baek}{Semelin
  et~al.}{2007}]{Semelin07}
Semelin B.,  Combes F.,   Baek S.,  2007, A\&A, 495, 389

\bibitem[\protect\citeauthoryear{{Shimabukuro} \& {Semelin}}{{Shimabukuro} \&
  {Semelin}}{2017}]{Shimabukuro17}
{Shimabukuro} H.,  {Semelin} B.,  2017, \mn@doi [MNRAS] {10.1093/mnras/stx734},
  \href {http://adsabs.harvard.edu/abs/2017MNRAS.468.3869S} {468, 3869}

\bibitem[\protect\citeauthoryear{{Vonlanthen}, {Semelin}, {Baek}  \&
  {Revaz}}{{Vonlanthen} et~al.}{2011}]{Vonlanthen11}
{Vonlanthen} P.,  {Semelin} B.,  {Baek} S.,   {Revaz} Y.,  2011, \mn@doi [A\&A]
  {10.1051/0004-6361/201116811}, \href
  {http://adsabs.harvard.edu/abs/2011A%26A...532A..97V} {532, A97+}

\bibitem[\protect\citeauthoryear{Wouthuysen}{Wouthuysen}{1952}]{Wouthuysen52}
Wouthuysen S.~A.,  1952, AJ, 57, 21

\bibitem[\protect\citeauthoryear{{Zahn}, {Mesinger}, {McQuinn}, {Trac}, {Cen}
  \& {Hernquist}}{{Zahn} et~al.}{2011}]{Zahn11}
{Zahn} O.,  {Mesinger} A.,  {McQuinn} M.,  {Trac} H.,  {Cen} R.,   {Hernquist}
  L.~E.,  2011, \mn@doi [MNRAS] {10.1111/j.1365-2966.2011.18439.x}, \href
  {http://adsabs.harvard.edu/abs/2011MNRAS.414..727Z} {414, 727}

\makeatother
\end{thebibliography}

\appendix

\end{document}